\documentclass{JINST}
\usepackage{graphicx}
\usepackage{epsfig}
\usepackage{url}
\usepackage{amssymb,fontenc,times,mathptmx}
\usepackage{wrapfig,rotating}
\usepackage[usenames]{color}
\makeatletter\let\default@color\current@color\makeatother
\title{
\begin{center}
Design and Electronics Commissioning of the Physics Prototype of a Si-W Electromagnetic Calorimeter for the International~Linear~Collider\\
\end{center}
}

\author{\centering 
\LARGE\bf The CALICE Collaboration
}

\author{\centering
J.Repond
\\ \it
Argonne National Laboratory,
9700 S.\ Cass Avenue,
Argonne, IL 60439-4815,
USA}

\author{\centering
J.Yu
\\ \it
University of Texas, Arlington, TX 76019, USA
}

\author{\centering
C.M.Hawkes, Y.Mikami, O.Miller, N.K.Watson,  J.A.Wilson
\\ \it
University of Birmingham,
School of Physics and Astronomy,
Edgbaston, Birmingham B15 2TT, UK
}

\author{\centering 
G.Mavromanolakis, M.A.Thomson, D.R.Ward, W.Yan
\\ \it
University of Cambridge, Cavendish Laboratory, J J Thomson Avenue, CB3 0HE, UK
}

\author{\centering
F.Badaud, D.Boumediene, C.C\^{a}rloganu, R.Cornat, P.Gay, Ph.Gris,
S.Manen, F.Morisseau, L.Royer
\\ \it
Laboratoire de Physique Corpusculaire de Clermont-Ferrand (LPC),
24 avenue des Landais,
63177 Aubi\`ere CEDEX, France
}

\author{\centering
G.C.Blazey, D.Chakraborty, A.Dyshkant, K.Francis, D.Hedin, G.Lima, V.Zutshi
\\ \it
NICADD, Northern  Illinois University,
Department of Physics,
DeKalb, IL 60115,
USA
}

\author{\centering 
J.-Y.Hostachy, L.Morin
\\ \it
Laboratoire de Physique Subatomique et de Cosmologie,
Universit\'{e} Joseph Fourier (Grenoble 1),
53, av. des Martyrs,
38026 Grenoble CEDEX, France
}

\author{\centering 
E.Garutti,
V.Korbel, 
F.Sefkow
\\ \it
DESY, Notkestrasse 85,
D-22603 Hamburg, Germany
}

\author{\centering  
M.Groll
\\ \it
Univ. Hamburg,
Physics Department,
Institut f\"ur Experimentalphysik,
Luruper Chaussee 149,
22761 Hamburg, Germany
}

\author{\centering 
G.Kim, D-W.Kim, K.Lee, S.Lee
\\ \it
Kangnung National University, HEP/PD, Kangnung, South Korea
}

\author{\centering 
K.Kawagoe, Y.Tamura
\\ \it
 Department of Physics, Kobe University, Kobe, 657-8501, Japan
}

\author{\centering 
D.A.Bowerman,
P.D.Dauncey, A.-M.Magnan, 
C.Noronha, % first ECAL paper only
H.Yilmaz, O.Zorba
\\ \it
Imperial College, Blackett Laboratory,
Department of Physics,
Prince Consort Road,
London SW7 2BW, UK 
}

\author{\centering 
V.Bartsch, 
J.M.Butterworth, % First ECAL paper only
M.Postranecky, M.Warren, M.Wing
\\ \it
Department of Physics and Astronomy, University College London,
Gower Street,
London WC1E 6BT, UK
}

\author{\centering 
M.Faucci Giannelli, 
M.G.Green, 
F.Salvatore, T.Wu
\\ \it
Royal Holloway University of London,
Dept. of Physics,
Egham, Surrey TW20 0EX, UK
}

\author{\centering 
D.Bailey, R.J.Barlow, 
M.Kelly, 
S.Snow, R.J.Thompson 
\\ \it
The University of Manchester, School of Physics and Astronomy,
Schuster Lab,
Manchester M13 9PL,
UK
}

\author{\centering 
M.Danilov, 
V.Kochetkov % ECAL paper only
\\ \it
Institute of Theoretical and Experimental Physics, B. Cheremushkinskaya ul. 25,
RU-117218 Moscow, Russia
}

\author{\centering 
N.Baranova, P.Ermolov, D.Karmanov, M.Korolev, M.Merkin,A.Voronin
\\ \it
M.V. Lomonosov Moscow State University (MSU),
Faculty of Physics,
Leninskiye Gory, Moscow, 119992, Russia
}

\author{\centering 
B.Bouquet, S.Callier, F.Dulucq, J.Fleury, H.Li,  G.Martin-Chassard, F.Richard,
Ch.de la Taille, R.Poeschl, L.Raux, M.Ruan,  N.Seguin-Moreau, F.Wicek, Z.Zhang
\\ \it
Laboratoire de L'acc\'elerateur Lin\'eaire,
Centre d'Orsay, Universit\'e de Paris-Sud XI,
BP 34, B\^atiment 200,
F-91898 Orsay CEDEX, France
}

\author{\centering 
M.Anduze, V.Boudry, J-C.Brient, C.Clerc, G.Gaycken, 
C.Jauffret, A.Karar,  P.Mora de Freitas, G.Musat, M.Reinhard, A.Roug\'{e},  A.L.Sanchez, J-Ch.Vanel, H.Videau
\\ \it
\'Ecole Polytechnique,
Laboratoire Leprince-Ringuet (LLR),
Route de Saclay,
F-91128 Palaiseau,
CEDEX France
}

\author{\centering 
J.Zacek 
\\ \it
Charles University, Institute of Particle \& Nuclear Physics,
V Holesovickach 2,
CZ-18000 Prague 8, Czech Republic  
}

\author{\centering 
J.Cvach, P.Gallus, M.Havranek, M.Janata, M.Marcisovsky, I.Polak, J.Popule, L.Tomasek, M.Tomasek, P.Ruzicka, P.Sicho, J. Smolik, V.Vrba, J.Zalesak 
\\ \it
Institute of Physics, Academy of Sciences of the Czech Republic, Na Slovance 2,
CZ-18221 Prague 8, Czech Republic
}

\author{\centering 
Yu.Arestov
\\ \it
Institute of High Energy Physics,
Moscow Region,
RU-142284 Protvino,
Russia
}

\author{\centering 
 A.Baird, R.N.Halsall
\\ \it
Rutherford Appleton Laboratory, Chilton, Didcot,
Oxon OX110QX, UK 
}

\author{\centering 
S.W.Nam, I.H.Park, J.Yang 
\\ \it
Ewha Womans University, Dept. of Physics,
Seoul 120,
South Korea
}

\abstract{The CALICE collaboration is studying the design of high performance electromagnetic and hadronic calorimeters for future International Linear Collider detectors. For the electromagnetic calorimeter, the current baseline choice is a high granularity sampling calorimeter with tungsten as absorber and silicon detectors as sensitive material. A ``physics prototype'' has been constructed, consisting of thirty sensitive layers. Each layer has an active area of $18 \times 18$\,cm$^{2}$ and a pad size of $1 \times 1$\,cm$^{2}$. The absorber thickness totals 24 radiation lengths. It has been exposed in 2006 and 2007 to electron and hadron beams at the DESY and CERN beam test facilities, using a wide range of beam energies and incidence angles. In this paper, the prototype and the data acquisition chain are described and a summary of the data taken in the 2006 beam tests is presented. The methods used to subtract the pedestals and calibrate the detector are detailed. The signal-over-noise ratio has been measured at $7.63 \pm 0.01$. Some electronics features have been observed; these lead to coherent noise and crosstalk between pads, and also crosstalk between sensitive and passive areas. The performance achieved in terms of uniformity and stability is presented.}

\preprint{draft v4.4}

\begin{document}

%\begin{frontmatter}



%\input{sections/Authors.tex}

% main text
%\section{}
%\label{}

\tableofcontents
%\linenumbers
%\pagewiselinenumbers

%\input{sections/introduction.tex}
%%\protect\newpage
\section{ECAL Si-W physics prototype}
\label{proto}

The CALICE (Calorimetry for the Linear Collider Experiments) collaboration~\cite{CALICE} is studying several designs of calorimeters for experiments at the future International Linear Collider (ILC)~\cite{ILC}. In order to have a mature design when the first results from the Large Hadron Collider experiments are expected and decisions on future colliders made, two main development phases have been defined. In the first phase, characterised by the construction of ``physics prototypes'', the feasibility of building a pixelised calorimeter is studied. In the second phase, ``technological prototypes'' will focus on the engineering details needed to insert such a calorimeter in an ILC detector. In both phases, the prototypes will be subject to extensive studies at beam test facilities. The data acquisition (DAQ) chain is common to all calorimeters, thereby easing their integration. The physics prototype for the electromagnetic calorimeter (ECAL) described in this document is a high granularity silicon-tungsten sampling calorimeter. Whereas the proof-of-principle of such detector has already been proven e.g. at LEP~\cite{SiCAL}~\cite{OPAL} and SLD~\cite{SLD}, the challenge for an ILC detector is the much larger size and the compactness required.

Physics at the ILC places stringent requirements on the detector~\cite{ILC}, including the ability to measure jet energies with a precision of at least $0.3/\sqrt{E_{\mathrm{jj}}/\mathrm{GeV}}$ \footnote{In terms of mass resolution, the requirement is $\sigma_{E_{j}}/E_{j} < 3.8$\,\%.}.
%, i.e. $\alpha(E_{j}) < 0.027 \times \sqrt{E_{\mathrm{jj}}/\mathrm{GeV}}$ with $\sigma_{E_{j}}/E_{j} =\alpha(E_{j})/\sqrt{E_{\mathrm{j}}/\mathrm{GeV}}$}. 
It has been demonstrated in Monte Carlo (MC) simulations~\cite{mark} that the Particle Flow Algorithm (PFA)~\cite{PFA} approach to the reconstruction of jet energies in terms of the individual particles can achieve this goal for typical jets at the ILC. The next step is to validate these MC simulations by testing real scale prototypes in beam tests. For an ILC detector, the ECAL has to be as compact as possible, in order to achieve the smallest possible effective Moli\`ere radius and to reduce the cost of the surrounding magnet. In addition, high transverse and longitudinal segmentation is required for the PFA to be effective.

\subsection{Physics considerations driving the design}
\label{protoPhysCons}

%As documented in the ILC Reference Design Report~\cite{ILC}, the physics potential of the ILC dictates the following basic requirements for a calorimeter for an ILC detector: to be as compact as possible, in order to achieve the smallest possible effective Moli\`ere radius and to reduce the cost of the surrounding magnet, and to be finely granulated, to allow a Particle Flow Algorithm (PFA)~\cite{PFA} approach for the reconstruction of jets in terms of the individual particle positions and energies.

With a sampling calorimeter approach, the choice of absorber material is driven by the need to separate particles in a jet and the need for a compact calorimeter. 
Separation of particles in the transverse direction requires an absorber material which gives narrow electromagnetic~(EM) showers and so needs to have a small Moli\`ere radius. Separation longitudinally requires a material with short EM showers, which pushes for a material with a small radiation length (X$_{0}$). The same requirement also results in a compact ECAL. 
%To keep the transverse and longitudinal dimensions of electromagnetic~(EM) showers as small as possible requires that the absorber material should have a small Moli\`ere radius, and a small radiation length (X$_{0}$). 
Tungsten is such a material, with a Moli\`ere radius of 9\,mm, and a radiation length of 3.5\,mm. Furthermore, the ratio of interaction length to radiation length is 27.4 so that hadronic showers typically develop later than EM showers. For the physics prototype, the pad size is $1 \times 1 $\,cm$^2$, comparable to the Moli\`ere radius. 
The choice of high resistivity silicon as the active material is motivated by the requirements of granularity and compactness. To contain high energy showers, the prototype should have around 24\,X$_{0}$ in total; this will ensure a containment of 99.5\,\% of the energy for 5\,$\mathrm{GeV}$ electron showers, and better than 98\,\% for 50\,$\mathrm{GeV}$ showers. Thirty layers were chosen to provide sufficient granularity. A small tungsten thickness in the first layers ensures a good energy resolution at low energy. 

%These requirements will ensure the construction of a representative part in depth of the final ECAL. 
%Futher studies~\cite{RDReport} have since shown that a pad size of $0.5 \times 0.5 $ cm$^2$ gives better performances, and this size has been chosen for the technological prototype, expected in 2008.

\subsection{Geometry and mechanical structure}
\label{protoGeom}
%\subsubsection{Description}
At normal incidence, the prototype has a total depth of 24~radiation lengths, achieved using 10~layers of 0.4\,X$_{0}$ (1.4\,mm) thick tungsten absorber plates, followed by 10~layers of 0.8\,X$_{0}$ (2.8\,mm) thick plates, and another 10~layers of 1.2\,X$_{0}$ (4.2\,mm) thick plates, with an overall thickness of 20\,cm. Each silicon layer has an active area of $18\times 18$\,cm$^{2}$, segmented into modules of $6\times 6$ readout pads of $1\times 1$\,cm$^2$ each. The active volume of the physics prototype therefore consists of 30~layers of $3\times 3$~modules, giving in total 9720~channels.

%\end{minipage}
%\hfill
%\begin{minipage}[r]{0.48\columnwidth}
The design and construction of the prototype presents a number of engineering challenges. A particular innovative effort has been made to minimise passive material zones and to keep the calorimeter as compact as possible, by incorporating half of the tungsten into alveolar composite structures.
%A particular innovative effort has been made to reduce the passive area whereby half of the tungsten is incorporated into alveolar composite structures. This state-of-the-art technology solution has of course an important impact, not only on the reduction of passive materials, but also on the compactness of this instrument. 
%The schematic 3D view of the prototype, shown in Figure~\ref{fig:3Dproto}, shows three independent structures, one for each thickness of tungsten. Each structure is fabricated by moulding preimpregnated carbon fibre (Cfi) and epoxy (``prepreg'') onto tungsten sheets, leaving free spaces between two layers to insert the detection units, called detector slabs.
Three independent structures can be distinguished, as shown in Figure~\ref{fig:3Dproto}, one for each thickness of tungsten. Each structure is fabricated by moulding preimpregnated carbon fibre (Cfi) and epoxy (``prepreg'') onto tungsten sheets, leaving free spaces between two layers to insert the detection units, called detector slabs.

\begin{figure}[h!]
\begin{minipage}[l]{0.45\columnwidth}
\centerline{\includegraphics[width=1.09\columnwidth]{./sections/figures/3DProtoH.eps3}}
\caption{\sl Schematic 3D view of the physics prototype.}
\label{fig:3Dproto}
%\end{figure}
\end{minipage}
\hfill
\begin{minipage}[l]{0.45\columnwidth}
%\begin{figure}[h!]%{l}{0.95\columnwidth}
\centerline{\includegraphics[width=1.09\columnwidth]{./sections/figures/Slab.eps3}}
\caption{\sl Schematic diagram showing the components of a detector slab.}
\label{fig:slab}
\end{minipage}
\end{figure}

One detector slab, shown in Figure~\ref{fig:slab}, consists of two active readout layers mounted on each side of an H-shaped supporting structure. The slab is shielded on both sides from the tungsten alveolar structure by an aluminium foil 0.1\,mm thick, to protect the silicon modules from electromagnetic noise and provide the wafer substrate ground. The H-shaped structure is 326\,mm long and 125.6\,mm wide, and has a mass of either 1.1, 2.2 or 3.3\,kg depending on the tungsten thickness.
The active layer is made of a 14-layer printed circuit board (PCB), 2.1\,mm thick and 600\,mm long, holding high resistivity silicon wafers 525\,$\mu$m thick (see Section~\ref{protoDescription}). The wafers are cut into square modules of $62\times 62$\,mm$^2$, separated from each other by a 0.15\,mm wide mounting gap.

Two detector slabs are inserted per layer, into the central and bottom cells of the alveolar structure (see Figure~\ref{fig:3Dproto}). The central and bottom slab active areas are formed by an array of $3 \times 2$ modules and a row of 3 modules respectively. To reduce overlapping passive areas, the two 
\begin{wrapfigure}{l}{0.60\columnwidth}
\centerline{\includegraphics[width=0.55\columnwidth]{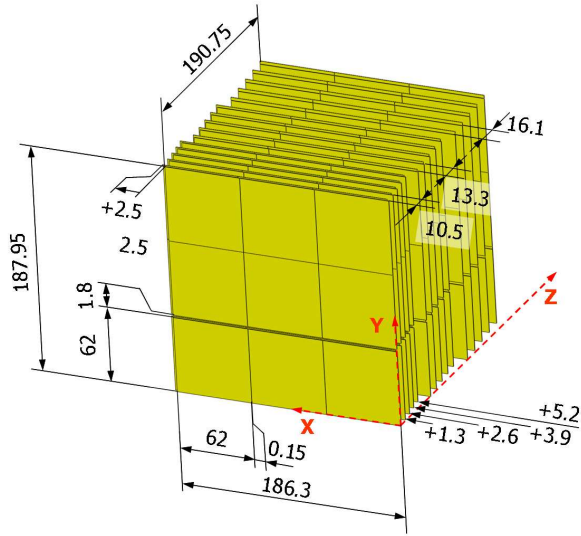}}
%\end{minipage}
\caption{\sl Details of the passive areas and layer offsets. Offsets are indicated by single-headed arrows. All distances are in mm.}
\label{fig:offsets}
\end{wrapfigure}
%\begin{figure}[h!]
detection layers of each slab are offset by 2.5\,mm in the $X$ direction, as shown in Figure~\ref{fig:slab}. Furthermore, the slabs in each substructure are offset by 1.3\,mm in the $X$ direction, as shown in Figure~\ref{fig:offsets}.

The passive area between modules is mainly due to two 1\,mm wide guard rings around the modules (see Section~\ref{protoDescription}). A larger passive area is located between the central and bottom slabs, and includes the two guard rings, two 0.15\,mm wide mounting gaps between module and PCB, two 0.3\,mm thick H structures, two 0.1\,mm thick aluminium shields, a 0.3\,mm wide global mounting gap, and a 0.4\,mm thick composite sheet, giving a total of 3.8\,mm.

\subsubsection{Fabrication processes}
\label{protoGeomGlue}
All the composite parts, i.e. H-shaped and alveolar structures, are made using 0.15\,mm thick carbon fibre and epoxy prepreg, TEXIPREG\textsuperscript{\textregistered} CC120 ET443~\cite{epoxypreg}, with an average thickness of 0.15\,mm.

Each alveolar structure is made in a single curing step of four hours at 135$^{\circ}$C. Fifteen metal cores are used to form the alveoli. They are wrapped with one layer of composite, and alternated with tungsten layers to obtain the final structure. Since the thermal expansion coefficient of carbon fibre is very close to that of tungsten, distortions during the curing are small. After curing, the metal cores are taken out, leaving empty spaces for the detector slabs (see Figure~\ref{fig:mould}). 

\begin{figure}[h!]%{l}{0.5\columnwidth}
\begin{minipage}[l]{0.45\columnwidth}
%\begin{wrapfigure}{l}{0.5\columnwidth}
\centerline{\includegraphics[width=0.98\columnwidth]{./sections/figures/AlveolaMould.eps3}}
\caption{\sl Moulding of an alveolar structure: metal cores alternate with tungsten, both wrapped in carbon fibre-epoxy composite.}
\label{fig:mould}
%\end{wrapfigure}
\end{minipage}
\hfill
\begin{minipage}[l]{0.45\columnwidth}
%\begin{wrapfigure}{l}{0.5\columnwidth}
\centerline{\includegraphics[width=0.98\columnwidth]{./sections/figures/WafGluing.eps3}}
\caption{\sl Gluing the modules; the accurate positioning of the modules is ensured by a grid of tungsten wires.}
\label{fig:wafGlu}
\end{minipage}
\end{figure}
%\end{wrapfigure}

The thin composite sheets located between two cells consist of three composite layers of thickness 0.4\,mm. Dimension and geometry tolerances of the structure are directly dependent on the machining of the mould: all pieces were ground with a resulting flatness of $\pm 0.01$\,mm. Metallic inserts, also included in the composite structure, are used to fasten each structure on its support table.

The mechanical and electrical connection of the silicon pads to the PCB is made using EPO-TEK\textsuperscript{\textregistered} 4110 conductive glue~\cite{glue}.
%To reduce the thickness of the detection layers, dots of glue provide the signal connections from the silicon to the PCB.
For the central (bottom) slabs, 216 (108) spots of conducting glue are deposited on each PCB by a pneumatically driven syringe dispenser, mounted on an X-Y automatic robot. The low viscosity nature of the gluing paste keeps the spots correctly shaped during the process. The pressure and the deposit time were chosen to obtain a final spot size between 3 and 4\,mm in diameter. The thickness of the glue is calibrated by using nylon spacers 0.12\,mm thick. Then, the six (three) modules for the central (bottom) slabs are put on the PCB by hand, as shown in Figure~\ref{fig:wafGlu}. 
%Précision positionnement Modules +-0.5mm : Cela correspond au jeu moyen entre la taille réelle du module (moyenne) 61.95 mm et la dimension de l’espace imposé par les fils de W (62.05)
The accurate positioning of each module is ensured during the polymerisation cycle by a grid of tungsten wires, holding the wafers in place. The expected accuracy is $\pm 0.05$\,mm, arising from the average module size of 61.95\,mm and the 62.05\,mm imposed by the grid of wires.

The curing process is fundamental to the performance of the connection. Curing parameters were chosen to have a good compromise between the final resistance of the connection ($<0.1\:\Omega$) and the mechanical stress applied on the spots of glue. The latter is due to the differential thermal expansion between the PCB and the silicon. The polymerisation of the glue was therefore done at $40 ^{\circ}$C for twelve hours.

%curing parameters were chosen to have a good compromise between the final resistance~($<0.1\:\Omega$), and the mechanical stress of the spot due to the differential thermal expansion between the PCB and the silicon. The polymerisation of the glue was therefore made at $40 ^{\circ}$C for twelve hours.

\subsubsection{Beam test configurations}

\begin{figure}[h!]
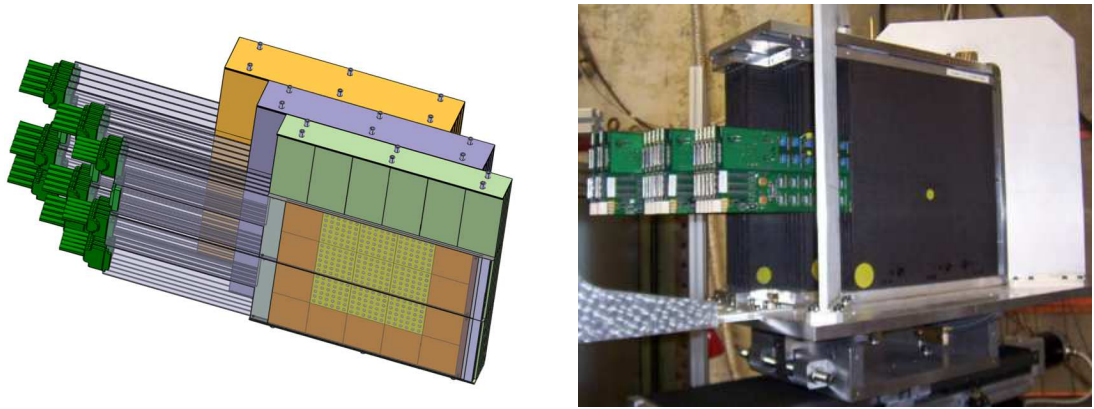

%\vspace*{-10mm}
\begin{minipage}[l]{0.5\columnwidth}
\centerline{\includegraphics[width=0.9\columnwidth]{./sections/figures/ProtoAngleSchema.eps3}}
%\caption{\sl Example of a rotated configuration, at the 45$^{\circ}$ angle setting.}
%\label{fig:AngleSchema}
\end{minipage}
\hfill
\begin{minipage}[r]{0.5\columnwidth}
%\end{figure}
%\begin{figure}[h!]%{l}{0.7\columnwidth}
\centerline{\includegraphics[width=0.9\columnwidth]{./sections/figures/ProtoAnglePict.eps3}}
\end{minipage}
\caption{\sl Example of a rotated configuration at the 45$^{\circ}$ angle setting: schematic 3D view (left) and picture (right) of the prototype at DESY in 2006 (note the staggered slabs and structures).}
\label{fig:AngleConf}
\end{figure}

To test the physics behaviour of the ECAL prototype with different impact angles of the beam, a rotational support was designed with six discrete angular configurations, from $0 ^{\circ}$ to $45 ^{\circ}$. The position of the three independent alveolar structures can be adjusted according to the chosen angle, to ensure the primary particle will travel through the centre of the active area, as shown in %Figures~\ref{fig:AngleSchema} and
Figure~\ref{fig:AngleConf}.
Each structure is mounted on rails, positioned manually and fixed using mechanical pins, with a precision of 0.5\,mm, giving an error of $\pm 0.035^{\circ}$ on the angle. The main error on the angle arises from the manual placement of the ECAL compared to the beam, and is $\pm 0.6^{\circ}$.
%\vspace*{-0.8cm}

In order to vary the impact position of the primary particle on the detector surface, the mechanical support of the prototype has been equipped with a control system allowing for a motion along both the $X$ ($\pm 150$\,mm) and $Y$ ($\pm 100$\,mm) axes, with a precision of $\pm 0.1$\,mm. The position can be piloted remotely during beam operation.

%To be able to choose the impact point of the beam on the prototype, a mechanical support has been designed with a control system allowing motion along both the X ($\pm 150$\,mm) and Y ($\pm 100$\,mm) axes, with a precision of $\pm 0.1$\,mm. The stage position can be moved remotely when the beam is on. 

\subsection{Description of the sensitive area}
\label{protoDescription}
%\subsubsection{Silicon module structure}

The sensors are made from four inch diameter silicon wafers. A resistivity greater than 5\,k$\Omega \cdot$\,cm is necessary to ensure a full depletion with a reasonable bias voltage, and to obtain low bulk leakage currents. A wafer thickness of 525\,$\mu$m was chosen, to obtain a signal-to-noise ratio of about 10 at the end of the whole readout electronics chain. The noise of the readout electronics results from the capacitance of the pad itself, the capacitance of the line on the PCB, and 
\begin{wrapfigure}{r}{0.5\columnwidth}%[h!]
%\begin{figure}[h!]
%\begin{minipage}[l]{0.5\columnwidth}
\centerline{\includegraphics[width=0.45\columnwidth]{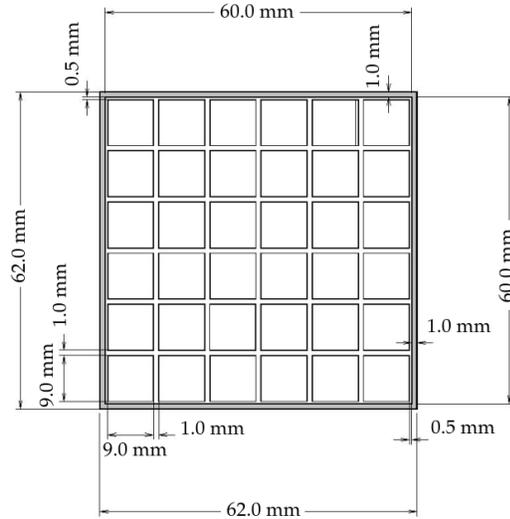}}
\caption{\sl Details of the matrix dimensions in a module (cathode side). The area shaded in grey corresponds to the guard ring.}
\label{fig:wafmat}
\end{wrapfigure}
%\end{minipage}
%\hfill
%\begin{minipage}[l]{0.5\columnwidth}
the preamplifier noise. A minimum ionising particle (MIP) produces about 80~electron-hole pairs per $\mu$m of silicon, hence 42,000~electrons are obtained for the 525\,$\mu$m thickness, giving an allowable noise range for the readout electronics of up to 4000~electrons. The crystallographic orientation of the silicon ($\langle 111 \rangle$) has only little importance; the detection of EM showers instead of single particles ensures that the fraction of energy lost through ``channeling effects'' is very small compared to the total signal.

Each wafer is used to make a module, consisting of a matrix of $6 \times 6$ PIN diodes (pads) of 1\,cm$^2$, as shown in Figure~\ref{fig:wafmat}. The overall dimensions take into account the guard rings. The anode side (N) is common to all pads. The PIN structure is made with ionic implantation. In order to keep the price and the rate of rejected processed modules low, the manufacturing must be as simple as possible, with a minimum number of steps during processing. The final detector will need about 3000\,m$^2$ of such detectors. %, necessary to limit the surface leakage currents.
The passivation layer of the processed module must be compatible with the gluing process described in Section~\ref{protoGeomGlue}. The main characteristics of a module are presented in Table~\ref{tab:wafmat}.

%\begin{table}[h!]
%\end{figure}
\begin{table}[h!]
\begin{center}
%\scriptsize{
\begin{tabular}{|l|c||l|c|}
\hline
Wafer diameter & 4 inch & Capacitance per pad & ~21\,pF\\
\hline
Wafer resistivity & 5\,k$\Omega \cdot$\,cm  & Full depletion bias voltage & ~150\,V\\
\hline
Wafer thickness & $525 \pm 16$\,$\mu$m & Nominal operating bias voltage & 200\,V \\
\hline
MIP deposit & 42 000 electrons & Break down voltage & $> 400$\,V \\
\hline
Average tile side & $61.95 \pm 0.05$\,mm & Leakage current at 200\,V & $< 350$\,nA (full matrix)\\
\hline
%Capacitance per pad & ~21\,pF\\
%\hline
%Full depletion bias voltage & ~150\,V\\
%\hline
%Nominal operating bias voltage & 200\,V\\
%\hline
%Break down voltage & $> 400$\,V\\
%\hline
%Leakage current at 200\,V & $< 300$\,nA (full matrix)\\
%\hline
\end{tabular}
%}
\caption{\sl Characteristics of a module.}
\label{tab:wafmat}
%\vspace{0.2cm}
\end{center}
\end{table}
%\end{minipage}
%\end{figure}

\subsubsection{Module test setup and results}

%\subsubsection*{Reception and test of the matrix}
Silicon PIN diode matrices are {\em a priori} very simple detectors. 
Having a whole ECAL covered with high resistivity silicon is however challenging, and implies the necessity to control the process at the earliest stage, in particular in terms of passivation. Therefore several manufacturers were used, enabling testing and validation of several technologies necessary to optimise the production in the future, and minimizing the cost.
%The challenge of having a whole ECAL covered with high resistivity silicon however implies a need to control the process at the earliest stage and so interact with several manufacturers, in particular in terms of passivation. Having several manufacturers enables testing and validation of several technologies necessary to optimise the production in the future, and also minimises the cost.

For the physics prototype, 270 modules are needed. The production of the modules extended from January 2003 to June 2007. All modules are made of raw wafers produced by Wacker~\cite{wacker}. The modules were produced by two manufacturers: the Institute of Nuclear Physics, Moscow State University~\cite{Moscow}, and ON Semiconductor Czech Republic~\cite{ONsemi} in conjunction with the Institute of Physics, Academy of Sciences of the Czech Republic, Prague.
%Modules from other producers are still under study, in particular for the technological prototype.

The two manufacturers use the same basic technique of ion implantation. Guard rings 1\,mm thick have been designed around the matrices to reduce surface leakage currents, with two different geometries.
%shown in Figure~\ref{fig:wafGR}. 
For both, the edge termination is made using the floating guard rings technique~\cite{JTE} which does not require any extra step in the fabrication process of modules, helping to reduce the overall cost.

%\begin{figure}[ht!]
%\centerline{\includegraphics[width=1.0\columnwidth]{./sections/figures/waferGR.eps}}
%\caption{\sl Corner view of the Czech (left) and Russian (right) modules.}
%\label{fig:wafGR}
%\end{figure}

\begin{figure}[h!]
\centerline{\includegraphics[width=0.9\columnwidth]{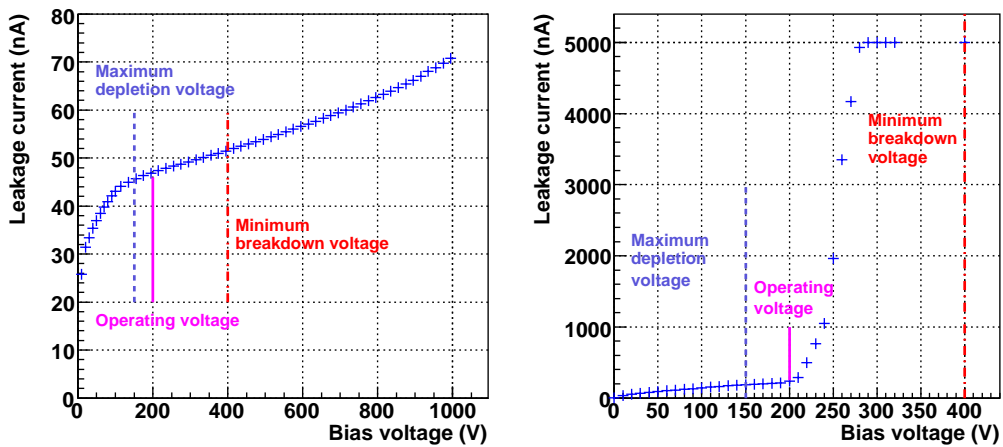}}
\caption{\sl Leakage current versus bias voltage. On the left is a good module, while on the right the breakdown voltage is too low.}
\label{fig:ivcurve}
\end{figure}

Before gluing, the modules are characterised one-by-one by measuring the leakage current as a function of bias voltage (I-V curve), the full depletion voltage (C-V curve) and the stability in time (leakage current versus time at nominal bias).
%In order to verify the characteristics of each matrix before gluing, a test bench has been designed to measure leakage current vs bias voltage (I-V curve) and full depletion voltage (C-V curve), as well as stability in time (leakage current versus time at nominal bias). The test bench consists of a high voltage power supply, a pico-ammeter and a C-V meter, as shown in Figure~\ref{fig:waftest}. Temperature and humidity are monitored.
If measurements are taken with a single biased pad, the values are found not to be representative of the performance of the module, especially in terms of breakdown voltage. 
All 36 pads are therefore biased, using a box specifically designed with 36 contacts made to the module using pin sensors mounted on springs, so as not to damage the surface. Two representative I-V curves taken with this setup are shown in Figure~\ref{fig:ivcurve}.

%\begin{figure}[h!]
%\centerline{\includegraphics[width=0.8\columnwidth]{./sections/figures/expSetup.eps}}
%\centerline{\includegraphics[width=0.8\columnwidth]{./sections/figures/biasBox.eps}}
%\caption{\sl Left: reception test bench for the silicon PIN diode matrix. Right: box specifically designed to bias all 36 pads of the tested module.}
%\label{fig:waftest}
%\end{figure}

% This issue explains most of the discrepancies observed between these measurements and those from the producer.
The test bench was used to characterise 550 modules, with an overall yield of 54\,\%. 
%The complete ECAL prototype contains 270 modules. 
Wafers were rejected if the full depletion voltage was greater than 150\,V, or if the breakdown voltage was less than 400\,V, or if the sum of the leakage currents of all 36 pads at the nominal operation voltage of 200\,V was greater than 350\,nA.

%The complete ECAL prototype contains 141 Russian and 129 Czech modules, which had a yield of 56\,\% and 52\,\% respectively. Russian modules were mainly rejected if the full depletion voltage was found to be greater than 150\,V, or the breakdown voltage less than 400\,V. Czech modules were mainly rejected if the leakage current was greater than 350\,nA at the nominal operation voltage of 200\,V.

\subsubsection{Calibration of the slabs}
\label{slabCalib}

\begin{figure}[h!]
\centerline{\includegraphics[width=1.0\columnwidth]{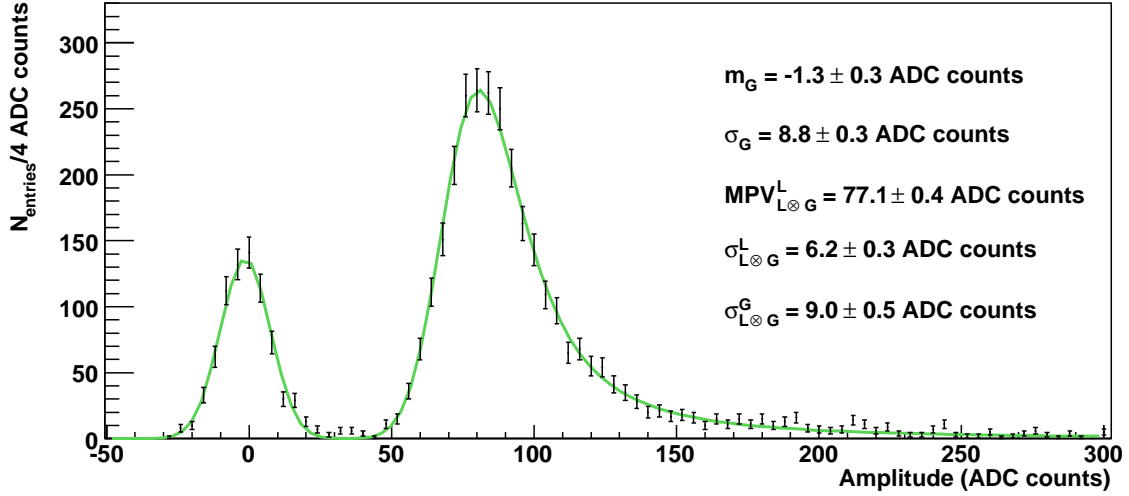}}
\caption{\sl Example of the cosmic-ray muon calibration of a half module showing the signal recorded by the corresponding readout chip. The first peak, centred on zero, is due to the electronics noise (pedestal) and is fitted with a Gaussian function of mean value $m_{G}$ and width $\sigma_{G}$. The second peak, centred on 82\,ADC~counts, corresponds to the MIP signal, and is fitted with a Landau (maximum probable value $MPV_{L \otimes G}^{L}$ and width $\sigma_{L \otimes G}^{L}$) convolved with a Gaussian distribution of width $\sigma_{L \otimes G}^{G}$. The overall fit function is defined as $Gauss(m_{G},\sigma_{G}) + Landau(MPV_{L \otimes G}^{L},\sigma_{L \otimes G}^{L}) \otimes Gauss(\sigma_{L \otimes G}^{G})$.}
\label{fig:cosmics}
\end{figure}

Once validated, the modules were assembled by gluing onto PCBs. Before assembling into a slab, the PCBs were tested in pairs using a cosmic-ray muon test bench, with a dedicated DAQ system. The gluing of each pad and the proper contact with the aluminum foil were checked and a first calibration of each half module, associated with a common readout chip (see Section~\ref{daqOnDet}), was possible. Initial MIP calibration results for each half module are obtained with a precision of approximately 5\,\%. An example is shown in Figure~\ref{fig:cosmics}.

%The results of the MIP calibration using the test bench are obtained per half matrix, as shown in Figure~\ref{fig:cosmics} for half a matrix, with an accuracy of around 5\,\%. They will be compared with a more accurate method in Section~\ref{calibtime}.

%Twelve hours of data taking is sufficient to obtain measurements of noise and MIP signal per half matrix, as shown in Figure~\ref{fig:cosmics}, with an accuracy of around 5\,\%. The results for the MIP calibration will be compared with a more accurate method in Section~\ref{calibTime}.

On average, the signal-to-noise ratio is measured to be $9.1 \pm 0.2$, with a standard deviation of $1.0 \pm 0.1$, and less than 0.1\,\% of the channels are identified as being dead. The quality of the points of glue can be estimated from the noise, since the measured noise increases with the resistivity of the contact point. The chemical passivation of some modules was found not to withstand the gluing process; it was found that the leakage current became greater than 10\,$\mu$A per module. This problem has now been solved by the producer.

%\subsubsection{Beam test observations}
The first slab was assembled in September 2004. Since that time, there has been no evidence of a degradation of the noise due to the gluing process.

%During the first period of beam test we met some problems in current consumption at the level of modules. After research we show that the problem came not from silicon but rather a bad isolation between via on the PCB and the carbon fibre structure. This problem was solving by putting a very thin isolated layer between the PCB and the carbon fibre structure.

%We also noted that when there is a large deposit of energy in a pad (> 100 PID) it can be observed from time to time on all the other pads of the matrix that the base line decreases proportionally with this energy. This phenomenon is not systematic and we have not yet find is origin. A possible explanation could be that there is time to time a bad contact between the aluminum foil and the matrix. In fact a common serial resistance of few ohms to all pad produce the same effect.

%\input{sections/squareEvents.tex}
\section{Electronics and data acquisition}
\label{daq}

The very-front-end (VFE) ASICs used to read out the silicon modules have been specifically designed for the prototype and are called FLC\_PHY3. The outputs of the VFE ASICs are transmitted to the off-detector electronics using differential analogue lines. The analogue-to-digital conversion is done on off-detector VME boards, using 16-bits ADCs. Each VME board can read out 96 VFE ASICs, and store the digitised data in memory for subsequent readout by an online software system written in C++.

\subsection{On-detector readout electronics}
\label{daqOnDet}

\subsubsection{FLC\_PHY3 ASIC}

%\begin{figure}[h!]
%\centerline{\includegraphics[width=0.95\columnwidth]{./sections/figures/AsicPicture.eps}}
%\caption{\sl Picture of FLC\_PHY3, packaged (left) and bare (middle), compared with a one euro coin (right).}
%\label{fig:phy3pic}
%\end{figure}

The FLC\_PHY3 VFE chip % (Figure~\ref{fig:phy3pic})
is an 18~channel, charge sensitive, front end circuit. It provides a shaped signal proportional to the input charge. Two chips are hence necessary to read out one module. Having two chips per module also introduces a redundancy into the system by decorrelating ASIC and module behaviour.
%ensures a decorrelation of ASIC and module behaviour, necessary for an accurate understanding of the whole system response.

\begin{wrapfigure}{l}{0.5\columnwidth}%[h!]
\centerline{\includegraphics[width=0.5\columnwidth]{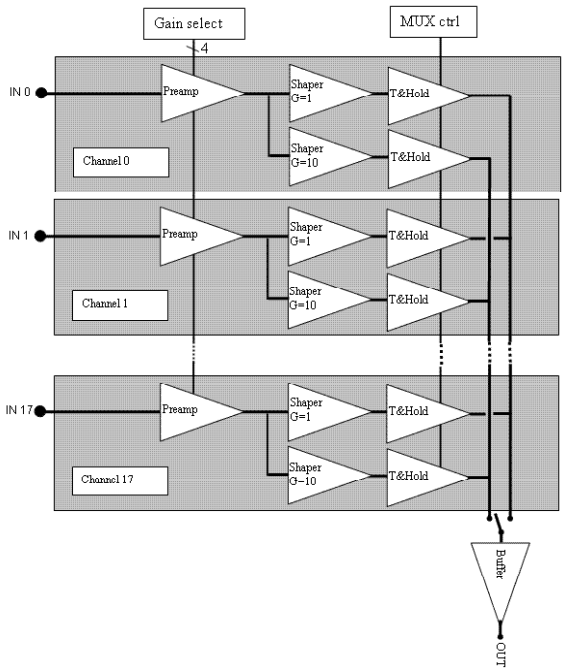}}
\caption{\sl General block schematic of FLC\_PHY3.}
\label{fig:block}
\end{wrapfigure}

Each channel is made of a variable gain charge preamplifier followed by two parallel shaping filters of different gain (1 and 10) using a CR-RC shaper, as shown in Figure~\ref{fig:block}. Each of these shapers is followed by a sample \& hold device (referred to as ``T\&Hold'' in Figure~\ref{fig:block}) driving a single multiplexed output. The bias of each stage is common to all channels.

The forward path of the preamplifier consists of an input PMOS transistor used in a common source configuration, ensuring a high transconductance. It is followed by a folded cascade bipolar NPN component. The last stage is made of a PMOS follower. The feedback is made with a switchable capacitance allowing a total feedback from 0.2\,pF to 3\,pF in steps of 0.2\,pF. The DC resistive feedback is ensured by a series of unbalanced current mirrors faking a 22.5\,M$\Omega$ resistor. This keeps the parallel noise negligible at the working frequency.

The shaper is made of a differential input and a single ended output amplifier with an open loop gain of around 100. The filter is a first order active pass-band with a frequency centred at around 1\,MHz and a peaking time of 200\,ns. This value has been chosen as a trade-off between the serial noise of the preamplifier which performs the optimal shaping around 1\,$\mu$s, the trigger latency which forbids a peaking time faster than 150\,ns, and the counting rates imposed by the ILC beam structure which require a recovery time below 300\,ns to avoid dead time on a pad.
The sample \& hold is ensured by a 2\,pF capacitance followed by a buffer using a Widlar structure.

\begin{figure}[h!]
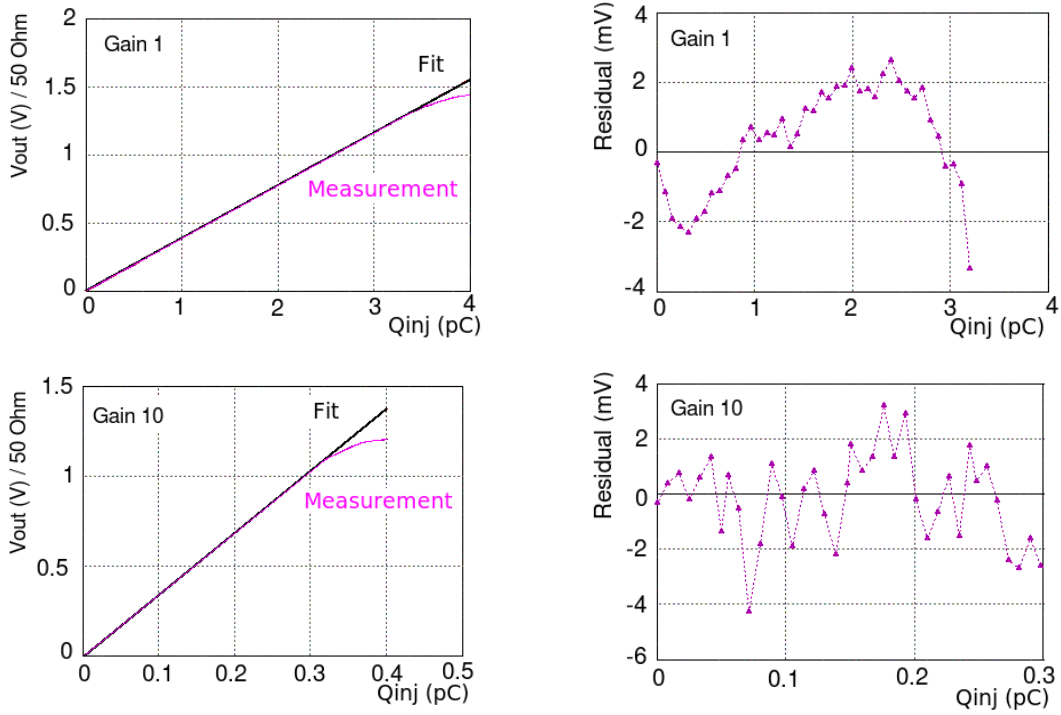

\begin{minipage}[l]{0.5\columnwidth}
\centerline{\includegraphics[width=0.85\columnwidth]{./sections/figures/ASIC_linearity_gain1.eps3}}
\end{minipage}
\hfill
\begin{minipage}[r]{0.5\columnwidth}
\centerline{\includegraphics[width=0.85\columnwidth]{./sections/figures/ASIC_residual_gain1.eps3}}
\end{minipage}
%\caption{\sl Output voltage of one channel of the FLC\_PHY3 chip as a function of the charge injected fitted by a linear function (top), and residual to the fit (bottom), for gain 1, for a feedback capacitance of 1.6\,pF.}
%\label{fig:linearityGain1}
%\end{figure}
\hfill
%\begin{figure}[h!]
\begin{minipage}[l]{0.5\columnwidth}
\centerline{\includegraphics[width=0.85\columnwidth]{./sections/figures/ASIC_linearity_gain10.eps3}}
\end{minipage}
\hfill
\begin{minipage}[r]{0.5\columnwidth}
\centerline{\includegraphics[width=0.85\columnwidth]{./sections/figures/ASIC_residual_gain10.eps3}}
\end{minipage}
\caption{\sl Output voltage \tt Vout \sl of one channel of the FLC\_PHY3 chip as a function of the charge injected \tt Qinj \sl fitted by a linear function (left), and residuals to the fit (right), for gain 1 (top) and gain 10 (bottom), for a feedback capacitance of 1.6\,pF.}
\label{fig:linearityGain1-10}
\end{figure}

The preamplifier gain is obtained with a nominal feedback capacitance of 1.6\,pF allowing a maximum input signal of around 4\,pC, equivalent to 600\,MIPs. A non-linearity of below 0.3\,\% up to 500\,MIPs is achieved, as shown in Figure~\ref{fig:linearityGain1-10}, and summarised in Table~\ref{tab:caract}. The residuals for the obvious non-linear regions above 3.27\,pC (0.3\,pC) for gain 1 (10) are not shown. In Monte Carlo (MC) simulation and beam test data of a 45\,$\mathrm{GeV}$ shower, the observed rate of hits recording more than 500 MIPs is about one hit per 500 events. In 100\,$\mathrm{GeV}$ shower however, the MC simulation predicts about 3.3 hits per event with energy deposit greater than 500 MIPs. This issue will be addressed in the next version of the chip \cite{JFleuryIEEE}.
%, in which case the non-linearities become important.

\begin{table}[h!]
\begin{center}
\begin{tabular}{|c|c|c|}
\hline
Shaper gain & 1 & 10\\
\hline
\tt{Qinj} max & 3.27\,pC & 0.33\,pC\\
\hline
\tt{Vout} max & 1.27\,V (over 50\,$\Omega$) & 1.12\,V (over 50\,$\Omega$)\\
\hline
Preamplifier Gain & 391\,mV$\cdot$\,pC$^{-1}$ & 3.44\,V$\cdot$\,pC$^{-1}$\\
\hline
%Measured Feedback & & \\
%capacitance       & \raisebox{1.5ex}[0pt]{1.27 pF} & \raisebox{1.5ex}[0pt]{1.45 pF}\\
Measured Feedback capacitance & 1.27\,pF & 1.45\,pF \\
\hline
Non-linearity & 0.32\,\% & 0.36\,\% \\
\hline
\end{tabular}
\caption{\sl Characterisation for a nominal feedback capacitance of 1.6\,pF.}
\label{tab:caract}
%\vspace{0.2cm}
\end{center}
\end{table}

 The equivalent charge for the noise has been measured on the VFE chip at around 4000~electrons, i.e. 0.1\,MIP, with an input capacitance of 80\,pF. This capacitance is made up of the detector capacitance of 20\,pF added to the PCB line capacitance of 60\,pF. The dynamic range hence goes from 0.1\,MIP to 600\,MIP, thus 13~bits using one shaper gain. The use of the two gains on the shaper in principle increases the dynamic range to 15~bits, although the high gain option was not used for the studies presented here.

From charge-injection measurements, the electronics crosstalk  between adjacent channels is found to be less than 0.1\,\%.

%has been measured for gain 1 and gain 10. A 3.4\,pC charge is injected in channel 4 for the Gain~1 crosstalk measurement, shown in Figure~\ref{fig:xtalk}. The average crosstalk is below 0.1\,\% in the 4 nearest channels to the hit channel.

%\begin{figure}[h!]
%\begin{minipage}[l]{0.3\columnwidth}
%\centerline{\includegraphics[width=0.98\columnwidth]{./sections/figures/xtalk_ch2.eps}}
%\end{minipage}
%\hfill
%\begin{minipage}[l]{0.3\columnwidth}
%\centerline{\includegraphics[width=0.98\columnwidth]{./sections/figures/xtalk_ch3.eps}}
%\end{minipage}
%\hfill
%\begin{minipage}[l]{0.3\columnwidth}
%\centerline{\includegraphics[width=0.98\columnwidth]{./sections/figures/xtalk_ch4.eps}}
%\end{minipage}
%\hfill
%\begin{minipage}[l]{0.3\columnwidth}
%\centerline{\includegraphics[width=0.98\columnwidth]{./sections/figures/xtalk_ch5.eps}}
%\end{minipage}
%\hfill
%\begin{minipage}[l]{0.3\columnwidth}
%\centerline{\includegraphics[width=0.98\columnwidth]{./sections/figures/xtalk_ch6.eps}}
%\end{minipage}
%\hfill
%\begin{minipage}[l]{0.3\columnwidth}
%\centerline{\includegraphics[width=0.98\columnwidth]{./sections/figures/xtalk_peak.eps}}
%\end{minipage}
%\caption{\sl Gain 1 crosstalk measurements, with a charge of 3.4\,pC injected in channel 4: output voltage from the hit channel (number 4) and its closest neighbours as a function of time. Bottom-right: signal magnitude (in mV) at peaking time vs channel number.}
%\label{fig:xtalk}
%\end{figure}

\subsubsection{Power consumption}
The consumption of the whole chip is 150\,mW corresponding to 8.3\,mW dissipation per channel. It would be easy to reduce that consumption by reducing the input transistor bias current, depending on the noise yield and the input capacitance. Setting a 200\,$\mu$A input current makes the consumption fall to 4\,mW per channel.

\subsubsection{Very Front End boards}
\label{daqvfe}

\begin{wrapfigure}{l}{0.5\columnwidth}%[h!]
\centerline{\includegraphics[width=0.5\columnwidth]{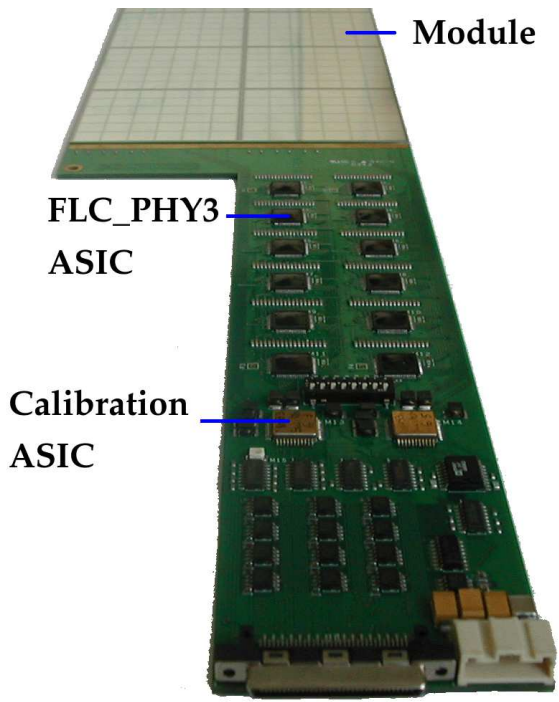}}
\caption{\sl Picture of the very front end PCB, and description of its elements. PCB thickness: 2.1\,mm, length: 600\,mm, largest width: 12.4\,mm.}
\label{fig:FEdescr}
\end{wrapfigure}
The 216 channels on the six modules of a PCB from a central slab (see Section~\ref{protoGeom}) are read out by twelve FLC\_PHY3 chips. Two 16-bit calibration ASICs, each having six channels, are also mounted on the PCB. The VFE ASICs are read out in parallel by the off-detector electronics using twelve differential analogue lines. To reduce the crosstalk below 0.1\,\%, the lines carrying the signals from the modules are spread on six different planes separated by ground planes, giving the 14-layer PCBs. Depending on the type of slabs on which they will be mounted (i.e. central or bottom row), three variants of the PCBs have been assembled: fully equipped with an array of $3 \times 2$ modules (30 used and 12 spares), as shown in Figure~\ref{fig:FEdescr}; and left or right equipped (compared to the orientation in Figure~\ref{fig:FEdescr}) with a row of three modules (15 used and 6 spares for each orientation).

A SCSI connector provides the interface to the off-detector electronics, allowing a full differential link in order to dissociate the digital ground in the readout crate from the detector ground. Each VFE ASIC has a separate channel to the connectors, i.e. there is no further multiplexing on the VFE board. The data transmission is analogue and the analogue-to-digital conversion is done on the off-detector board. A differential buffer is used to convert the single-ended signal provided by the VFE ASIC before transmission. A reference is used for each chip. The digital links are RS-422, allowing a wide common mode range to correct the voltage shift between the VFE PCB, supplied with a negative voltage to optimise the noise, and the off-detector board, supplied with a positive voltage.

\subsection{Off-detector electronics}
\label{daqOffDet}
%\subsubsection{Requirements}
%\label{daqReq}
The basic requirements of the off-detector electronics are: to distribute the sample \& hold signal required by the VFE electronics within a latency of 180\,ns and with an uncertainty of less than 10\,ns; to provide the digital sequencing necessary to multiplex the analogue signals from the VFE PCBs; to digitise the signals; and to store the data. The analogue signals have a 13-bit dynamic range, and the electronics is required not to contribute significantly to the analogue noise. Assuming a standard beam test spill structure, the target for the electronics is to run at an event rate of 1\,kHz during the beam spill, with an overall average rate of 100\,Hz.

\subsubsection{Implementation}

The calorimeter readout is based on the ``Calice Readout Cards''
(CRC)~\cite{warren}. These are custom-designed, 9U, VME boards derived from
the Compact Muon Solenoid tracker Front End Driver readout
boards~\cite{fed} but with major modifications to the readout and
digitisation sections.

Each CRC consists of eight front-end (FE) sections feeding into a single
back-end (BE) which provides the interface to VME. The whole board is clocked
from an on-board 40\,MHz 
oscillator.

Each FE section can control one full or two half VFE PCBs, totalling 12 VFE
readout ASICs, each with 18 multiplexed channels, and hence a total of 216
channels. The FE section consists of a pair of 68-pin mini-SCSI connectors,
twelve 16-bit ADCs (one for each readout ASIC) and a Xilinx Virtex-II FPGA.
Mini-SCSI cables of 10\,m length
are used to connect the FEs directly to the 
VFE PCBs. The digital signals on the mini-SCSI connector are all LVDS and are
tracked directly to the FPGA.
All VFE ASIC
control and clocking signals are generated in the FE FPGA, including the
time-critical sample \& hold signal edge. The FPGA derives a 160\,MHz 
clock from the 40\,MHz board clock, allowing signal timing to 6.25\,ns.
The
FE section also contains two 16-bit DACs for calibration purposes. These can
be internally fed back into the ADCs directly, or can be sent onto the
mini-SCSI connector to allow calibration of the VFE ASIC.
The total data volume within each FE per
event for 18 multiplex samples is 512 bytes, including an
eight-byte 
header.

The BE section consists of a Xilinx Virtex-II FPGA controlling an
8\,MByte memory. It distributes configuration data to each FE section and
gathers the digitised event data from each before storing it in the memory
for subsequent readout.

Each CRC is capable of reading out 1728 channels. The data volume per event
from these channels is 4\,kBytes, allowing 2000 events to be stored in the
memory before readout is required. Six CRCs are required for the full ECAL
readout of 9720 channels and these are housed in a single VME 
crate. The VME interface used is the SBS620 VMEbus-PCI bridge~\cite{sbs}.

Some systems other than the CRC VME readout are also integrated into
the online system. Tracking (see Section~\ref{beamtests}) is done using commercial VME TDC units
which are installed in spare slots of the CRC crates. Slow data control
and readout is implemented via sockets to independent slow control
systems. All slow data are written into the data files to ensure
the raw data are self-contained.

\subsubsection{Trigger and timing}
A specific VME slot near the centre of the crate is chosen as the
trigger control and distribution CRC and a custom backplane is
used to take
triggers from this central location to the other CRCs in the crate.
The backplane is a simple point-to-point PCB, press-fit to
the J0 connectors. The trigger distribution and control is
generated in the BE FPGA of the trigger slot CRC and fanned out
to the other CRCs. A second fanout on the CRC allows a duplicate
set of trigger signals to be taken via a cable to an identical
backplane in a second crate, for readout of more than one calorimeter.

The trigger input signals are LVDS and are fed into the J2 connector.
The trigger logic is performed within the BE FPGA, allowing 
significant complexity in the trigger condition to be set via
software. This also controls the trigger busy signal, preventing
further triggers until the digitisation of the VFE analogue data
has been completed.

The trigger
rising edge provides the system synchronisation signal and all timings,
including those for the time-critical sample \& hold, are derived from this edge.
To ensure complete events can be built from the same trigger,
independent trigger counters are implemented in each of the
FE and BE FPGAs, and are read periodically, both during and at the
end of each beam spill.
If any discrepancy between the counters
is observed, all data from the spill following the last valid counter readout are
discarded offline.

All trigger inputs, as well as several other input signals, are
recorded in a trigger history buffer. This consists of 32 bits,
one for each signal, sampled on the 40\,MHz clock, recording their
state for 256 samples around the trigger time. This allows the
detailed time structure of the trigger logic to be observed as
well as the readout of several other beam line elements such as
veto scintillator counters and a \v{C}erenkov detector.

\subsubsection{Performance}
The complete system performed well in the beam test environment.
The sample \& hold is distributed on the derived 160\,MHz clock
within a minimum of 160\,ns, allowing the rest of the latency of
180\,ns to be implemented as a software-specifiable
delay in the FE FPGA. The
jitter on this is dominated by rounding to the 6.25\,ns clock period. 
The 16-bit ADCs allow both positive and negative signals
although only the upper half is used for the ECAL readout. This
still allows them to meet the 13-bit dynamic range requirement, with the
two extra bits covering any differential non-linearities or other 
biases in the ADCs. 
The CRC noise when the ECAL is disconnected is very low, with
an average of 1.4\,ADC~counts, compared to around 5.9\,ADC~counts from
the ECAL VFE electronics (see Section~\ref{noiseUnif}). 
The trigger counter synchronisation is very reliable, with a
rate of less than one trigger missed on any CRC per million events, resulting
in an average event loss of less than one per thousand.

Two significant problems were identified. Firstly, insertion and removal of the cables from the mini-SCSI connectors resulted in breaks in the solder joints, in nearby traces, and in solder connections to nearby FE components, most of which have been fixed by hand. Secondly, there were also problems with the reliability of some of the PCB layer interconnects, and several wires have been added to bypass vias which developed high resistivity. Overall, the system had errors on 3 out of 112 FE sections, none of which were used for data taking.

\subsubsection{Readout software}
The online software must support the electronics requirements of 
1\,kHz instantaneous and 100\,Hz average event rates. 
The maximum event size to be handled is 64\,kBytes, allowing for
readout of two calorimeters and other beam line equipment.
The system is
also required to be flexible, so as to handle different beam lines 
and spill structures.

The code is written in C++ and is specific to Linux, being POSIX
compliant as much as possible. 
No database is used in the online system; all configuration of
the hardware is stored in the run data files so that any run can
be fully characterised from the data file itself.

The full system achieves a readout rate of 120\,Hz when not limited
by the beam intensity or spill structure. The average rate during
higher intensity spills is found to be around 90\,Hz. The trigger
rate during spills does not reach 1\,kHz but is limited to around
600\,Hz. This is mainly determined by the fraction of events for which the
trigger counters are read out. These are read
with a high frequency to be conservative in checking for
synchronisation errors; a lower fraction would allow the specified
trigger rate of 1\,kHz to be achieved.

The system has recorded up to 10 million events per day and has run efficiently 
for several beam periods totalling nearly three months. Including 2007 data, around 300 million beam events have been taken, amounting to around 30\,TBytes of data.

\section{Beam tests}
\label{beamtests}

%\subsection{Overall objectives of the beam tests}
%\label{beamtestObjs}

Large scale beam tests are conducted in order to demonstrate the principle of a highly granular Si-W electromagnetic calorimeter. Data analyses are ongoing which will quantify the energy and spatial resolutions, and validate the existing models of electromagnetic and hadronic showers, e.g. the various GEANT4~\cite{G4} models. The studies in this article focus on the basic hardware performance when the device is exposed to high energy particles.

%The beam tests have three primary objectives. The first is to characterise the physics prototype. In order to prove the validity of the concept for a calorimeter for an ILC detector, linearity in energy and a good energy resolution must be proven over a large range of energies. Position and angular resolutions need to be measured across the detector, and the uniformity of the wafer response needs to be proven. The shower also needs to be characterised, by studying the longitudinal and transverse shower profiles. The second objective is to validate the simulation, performed using the GEANT4~\cite{G4} application Mokka~\cite{Mokka}, against a realistic detector. Once the simulation is trusted and realistic, full detector studies can be done to optimise the detector. The third objective is to identify hardware problems, by studying the sensors and the electronics behaviour in real electromagnetic showers. Only the last objective is presented in this paper; the two others will be treated in subsequent publications.

\subsection{Beam lines at DESY and CERN in 2006}
\label{beamtestDescr}

\begin{figure}[h!]
\begin{center}
\centerline{\includegraphics[width=0.75\columnwidth]{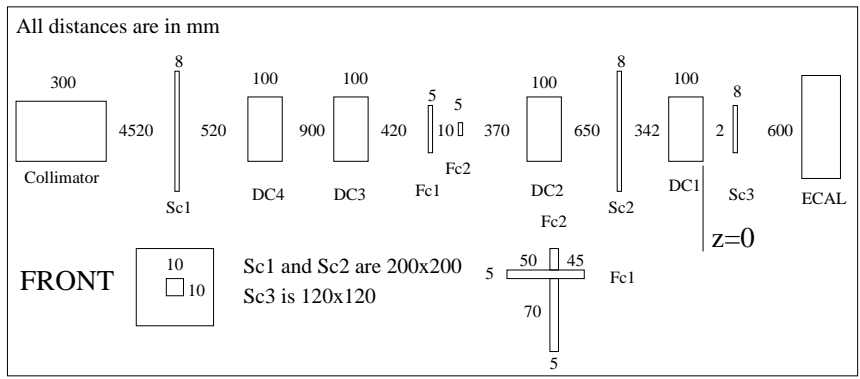}}
\caption{\sl Sketch of the DESY beam test setup in May 2006. All distances and components dimensions are in mm, along the beam axis. The beam enters from the left side. 
See text for explanations of the components.}
\label{fig:Desy_setup}
\end{center}
\end{figure}

In May 2006, the ECAL prototype was tested with electron beams at DESY, equipped only with 24 layers of central slabs (see Section~\ref{protoGeom}). A sketch of the beam test setup is shown in Figure~\ref{fig:Desy_setup}.
The beam trigger is defined by the coincidence signal of two of the three scintillator counters, referred to as Sc1, 
Sc2 and Sc3 in Figure~\ref{fig:Desy_setup}. The size of the beam can also be defined by two scintillator counters, mounted in a cross shape (Fc1 and Fc2 in Figure~\ref{fig:Desy_setup}). Four drift chambers (DC1, DC2, DC3 and DC4) are used to monitor the beam and reconstruct tracks.

\begin{figure}[h!]
\begin{center}
\centerline{\includegraphics[width=0.85\columnwidth]{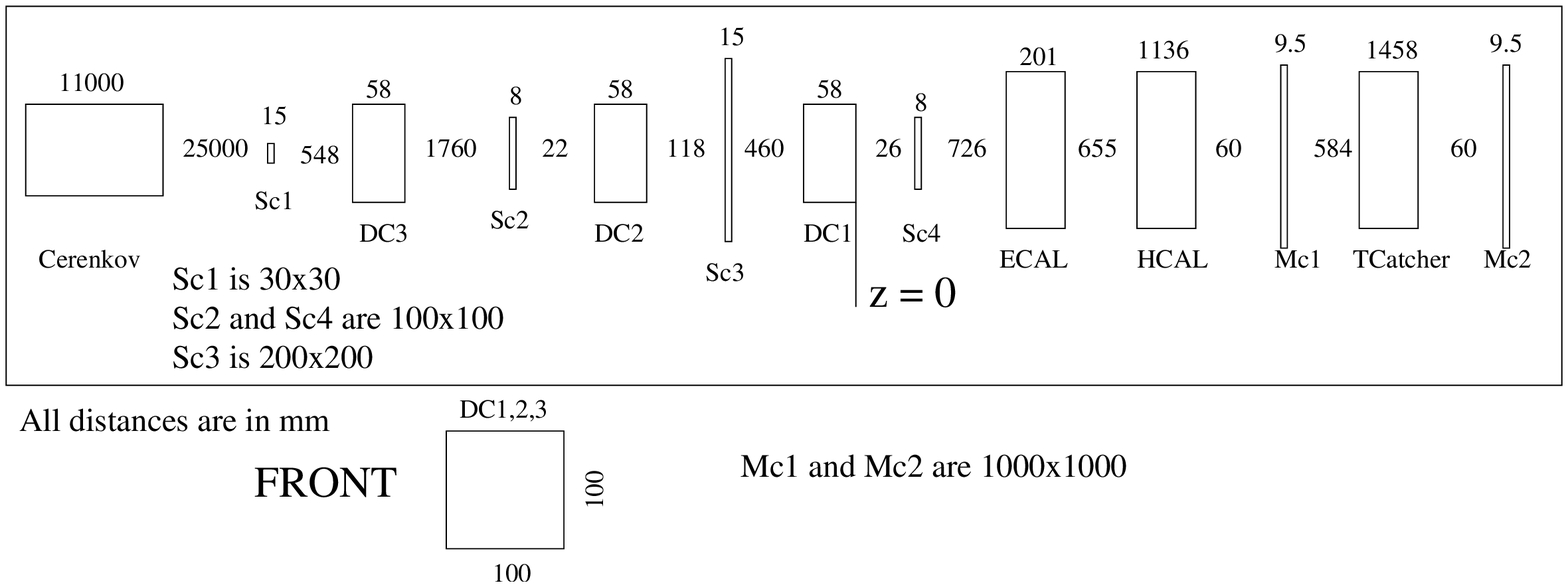}}
\centerline{\includegraphics[width=0.85\columnwidth]{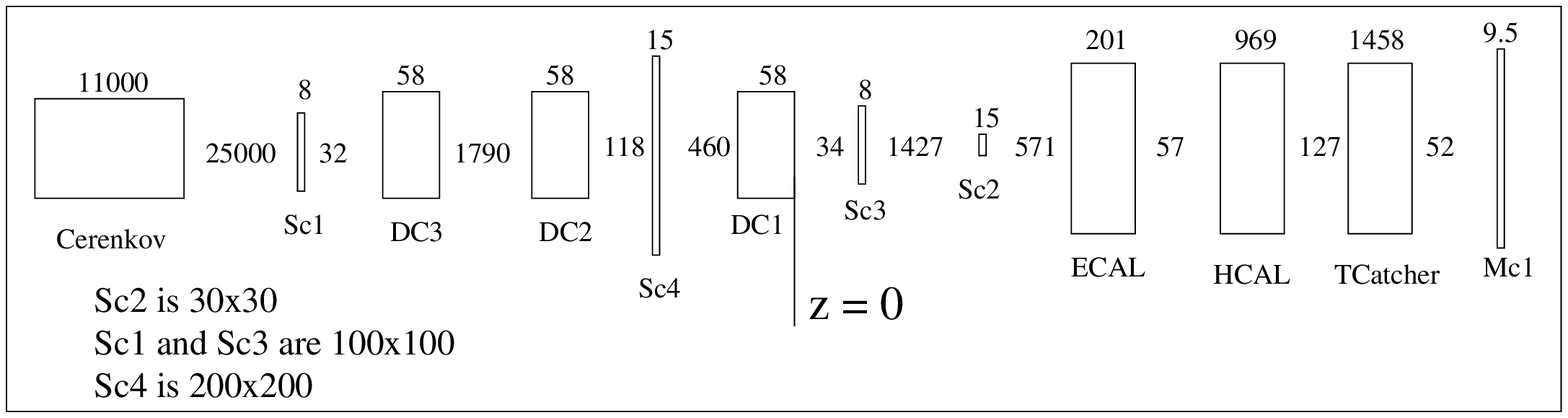}}
\caption{\sl Sketch of the CERN beam test setups in August (top) and October (bottom) 2006. All distances and components dimensions are in mm, along the beam axis. The beam enters from the left side.}
\label{fig:CERN_setup}
\end{center}
\end{figure}

%The resulting $e^-$ beam spot is {\bf xx.x} cm. 
In August and October 2006, the ECAL was tested at the H6 area at CERN, in combination with an analogue HCAL prototype~\cite{AHCAL} and a Tail Catcher and Muon Tracker prototype~\cite{TCMT}, using electron and pion beams. The 30 layers of the ECAL prototype were equipped with central slabs. Sketches of the beam line for both periods are presented in Figure~\ref{fig:CERN_setup}. In August, the distance between ECAL and HCAL was large enough to allow the ECAL to be rotated. In October however, the accent was put on minimising this distance so as to use the ECAL as a tracker for pion events.
The beam trigger is defined by the coincidence signal of two of the three scintillator counters (Sc2, Sc3 and Sc4 in the August setup, Sc1, Sc3 and Sc4 in the October setup).
When it was not used in the trigger, the largest scintillator (Sc3 in August, Sc4 in October) was still read out and so could be used as a beam halo veto offline, using a lack of coincidence between the $10 \times 10$ and $20 \times 20$\,cm$^2$ scintillators.
Three delay wire chambers~\cite{DCCern} (DC1, DC2, DC3 in Figure~\ref{fig:CERN_setup}) are used to monitor the beam and reconstruct tracks.
A \v{C}erenkov detector is used to separate electrons from pions.   
Muons are identified by two scintillators (Mc1, Mc2) placed behind the HCAL prototype.
%\vspace*{1cm}

\subsection{Summary of the data collected}
\label{beamtestSum}

Various scans were performed in order to characterise the prototype. The energy of the electron beams was varied from 1 to 6\,$\mathrm{GeV}$ at DESY, and from 6 to 50\,$\mathrm{GeV}$ at CERN, to study linearity and resolution. The beam was positioned near the centre, edge and corner of the wafers to study the uniformity of the wafers and the impact of the passive areas. Events with positrons were also taken from 6 to 50\,$\mathrm{GeV}$. The ECAL prototype was rotated by 10$^{\circ}$, 20$^{\circ}$, 30$^{\circ}$ and 45$^{\circ}$ with respect to the beam direction to study the effects of angular incidence on the linearity and resolution.
Finally, $\pi^+$ and $\pi^-$ beams were taken with energies from 6 to 80\,$\mathrm{GeV}$.

\begin{figure}[h!]%{l}{0.6\columnwidth}%[h!]
\centerline{\includegraphics[width=0.75\columnwidth]{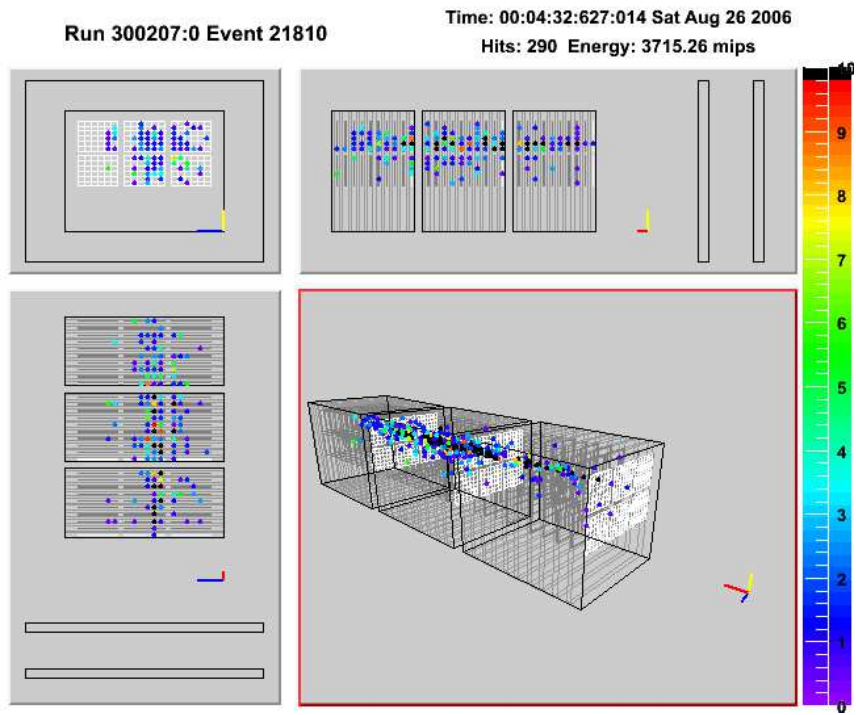}}
\caption{\sl Event display showing a 30\,$\mathrm{GeV}$ electron showering in the ECAL, in a run taken at CERN in August 2006. Only hits with an energy above 0.5\,MIP are displayed. The colour scale represents the energy deposited per pad, in units of MIP. Top-left: $X-Y$ projection, top-right: $Y-Z$ projection, bottom-left: $X-Z$ projection.}
\label{fig:eDisp}
\end{figure}

Over 22~million events were collected with electron beams, as detailed in Table~\ref{tab:Data_summary}, and 25~million events with pion beams. In addition, more than 57 million muon events were collected for calibration purposes (see Section~\ref{calib}).
An event display of an electron event at 30\,$\mathrm{GeV}$ is shown in Figure~\ref{fig:eDisp}.
\begin{table}[h!]
\begin{center}
\scriptsize{
\begin{tabular}{|c||c|c|c|c|c||c|c|c|c|c|c|} \hline
Location       & \multicolumn{5}{c||}{DESY (kEvt)} & Location       & \multicolumn{5}{|c|}{CERN (kEvt)} \\
\hline
Angles         & 0 $^{\circ}$      & 10 $^{\circ}$     & 20 $^{\circ}$      & 30 $^{\circ}$      & 45 $^{\circ}$    &  Angles        & \multicolumn{2}{c|}{0 $^{\circ}$}  & 20 $^{\circ}$   & 30 $^{\circ}$ & 45 $^{\circ}$ \\
\hline
 Particle     & \multicolumn{5}{c||}{e$^-$} & Particle  & e$^+$ & e$^-$ & \multicolumn{3}{c|}{e$^-$} \\
%\hline
%Particle      & \multicolumn{5}{c||}{e$^-$} & & e$^+$ & e$^-$  & e$^-$ & e$^-$ & e$^-$\\
\hline \hline
1~$\mathrm{GeV}$          & 400    & 300    & 345     & 200     & 200  & 6~$\mathrm{GeV}$ & 208    & 128  &        &        &      \\
1.5~$\mathrm{GeV}$        & 486    & 200    & 200     & 300     & 200  & 8~$\mathrm{GeV}$ &        & 218  &        &        &       \\
2~$\mathrm{GeV}$          & 400    & 200    & 200     & 300     & 200 & 10~$\mathrm{GeV}$ & 152    & 469  &   112  &  594   &   530\\
3~$\mathrm{GeV}$          & 304    & 200    & 200     & 324     & 200 & 12~$\mathrm{GeV}$ &        & 211  &        &        &      \\
4~$\mathrm{GeV}$          & 400    & 224    & 200     & 300     & 200 & 15~$\mathrm{GeV}$ & 476    & 325  &        &  181   &   244\\
5~$\mathrm{GeV}$          & 304    & 300    & 200     & 325     & 200 & 16~$\mathrm{GeV}$ & 310    &      &        &        &      \\
6~$\mathrm{GeV}$          & 594    & 688    & 200     & 185     & 200 & 18~$\mathrm{GeV}$ & 303    & 231  &        &        &      \\
\cline{1-6}
\cline{1-6}
Total          & 2888   & 2112   & 1545    & 1934    & 1400 & 20~$\mathrm{GeV}$ & 390    & 590  &   110  &  330   &   208   \\
\cline{1-6}
\multicolumn{6}{c||}{}                                      & 30~$\mathrm{GeV}$ & 409    & 685  &   270  &  550   &   531   \\
\multicolumn{6}{c||}{}                                      & 40~$\mathrm{GeV}$ &        & 347  &        &  280   &   311   \\
\multicolumn{6}{c||}{}                                      & 45~$\mathrm{GeV}$ &        & 933  &   250  &  753   &   551   \\
\multicolumn{6}{c||}{}                                      & 50~$\mathrm{GeV}$ & 305    &      &        &        &         \\
\cline{7-12} \cline{7-12}
\multicolumn{6}{c||}{} & total   & 2553 & 4137 & 742 & 2688 & 2375 \\
\cline{7-12}
\end{tabular}
}
\caption{\sl Summary of the 22.4 million triggers collected at DESY and CERN with electron beams.}
%\vspace{0.2cm}
\label{tab:Data_summary}
\end{center}
\end{table}

%\input{sections/titles.tex}
%\section{Commissioning of the ECAL prototype}
\section{Pedestal, noise, and crosstalk around module edge}
\label{pednoise}

A data-taking run consists of 500~pedestal events, 500~events with charge injection via the calibration chips, and between 10,000 and 40,000~beam data events, after which the sequence repeats. The pedestal and calibration events are taken with a random trigger occurring outside the beam spill period. In addition, pedestal events are taken during the beam spill (every 25 beam events on average), to check the influence of the beam on the pedestals. The determination of the pedestals and noise for the individual channels is described in the following sections. In these sections, the conversion from ADC counts to MIP equivalent units uses the factor 46\,ADC~counts per MIP (see Section~\ref{calib}).

%\subsection{Pedestal and noise measurements}
\subsection{Pedestal subtraction procedure}
\label{pedestal}

For a given channel, the pedestal is defined by the mean value of the signal recorded with no beam, i.e. coming from electronics, and the noise by the corresponding standard deviation. Instabilities occurring in the time between two sets of pedestal events need to be considered and accounted for in the pedestal subtraction procedure described here. Correlations in the noise also need to be checked, and accounted for if induced by pedestal drifts, and will be described in Section~\ref{sips}. Once the pedestal subtraction procedure is trusted, the uniformity, the stability in time of any remaining offsets, and the noise in beam-induced events all need to be measured. The results are presented in Section~\ref{pednoiseUnif}.

\subsubsection{VFE PCB-wise correlated pedestal shifts}
\label{pedsubtr}

\begin{figure}[h!]
\centerline{\includegraphics[width=0.6\columnwidth]{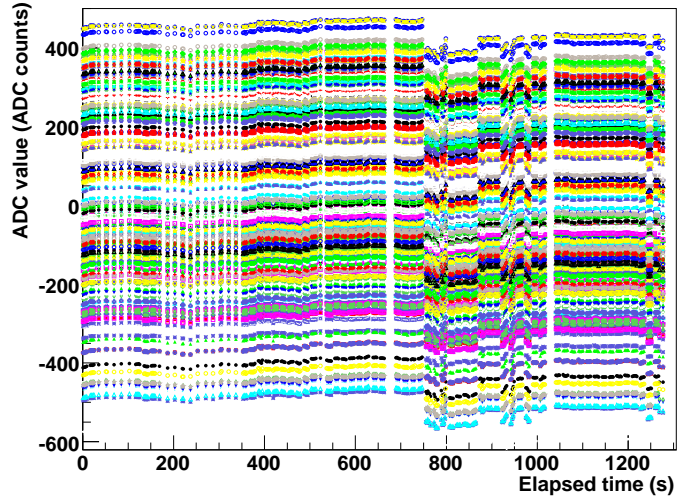}}
%\end{minipage}
\caption{\sl ADC values recorded in all 216 channels (each set of points represents one channel) of the VFE PCB situated in the 24$^{th}$ layer, as a function of time (seconds since the beginning of the run), in beam events in a muon run taken at CERN in August 2006.}
\label{fig:pedvstime}
\end{figure}

 The pedestals of all the channels of one VFE PCB are shown, as a function of time, in Figure~\ref{fig:pedvstime}. In this figure, a clear shift in the raw ADC values can be seen between two sets of pedestal events (identified by the blank columns at 680\,s and 1030\,s). This unexpected effect of pedestal shifts of up to 100\,ADC counts (equivalent to approximately two MIPS) for all the channels of a VFE PCB is observed frequently in beam data events. The origin of these shifts has been traced to the non-isolation of the VFE PCB power supply lines, which results in changes to the working point of the output signal lines. This will be corrected in the next PCB design. 

%\begin{wrapfigure}{l}{0.5\columnwidth}%[h!]
%\begin{minipage}[l]{0.48\columnwidth}

\begin{figure}[h!]
\begin{minipage}{0.5\columnwidth}
\centerline{\includegraphics[angle=-90,width=0.98\columnwidth]{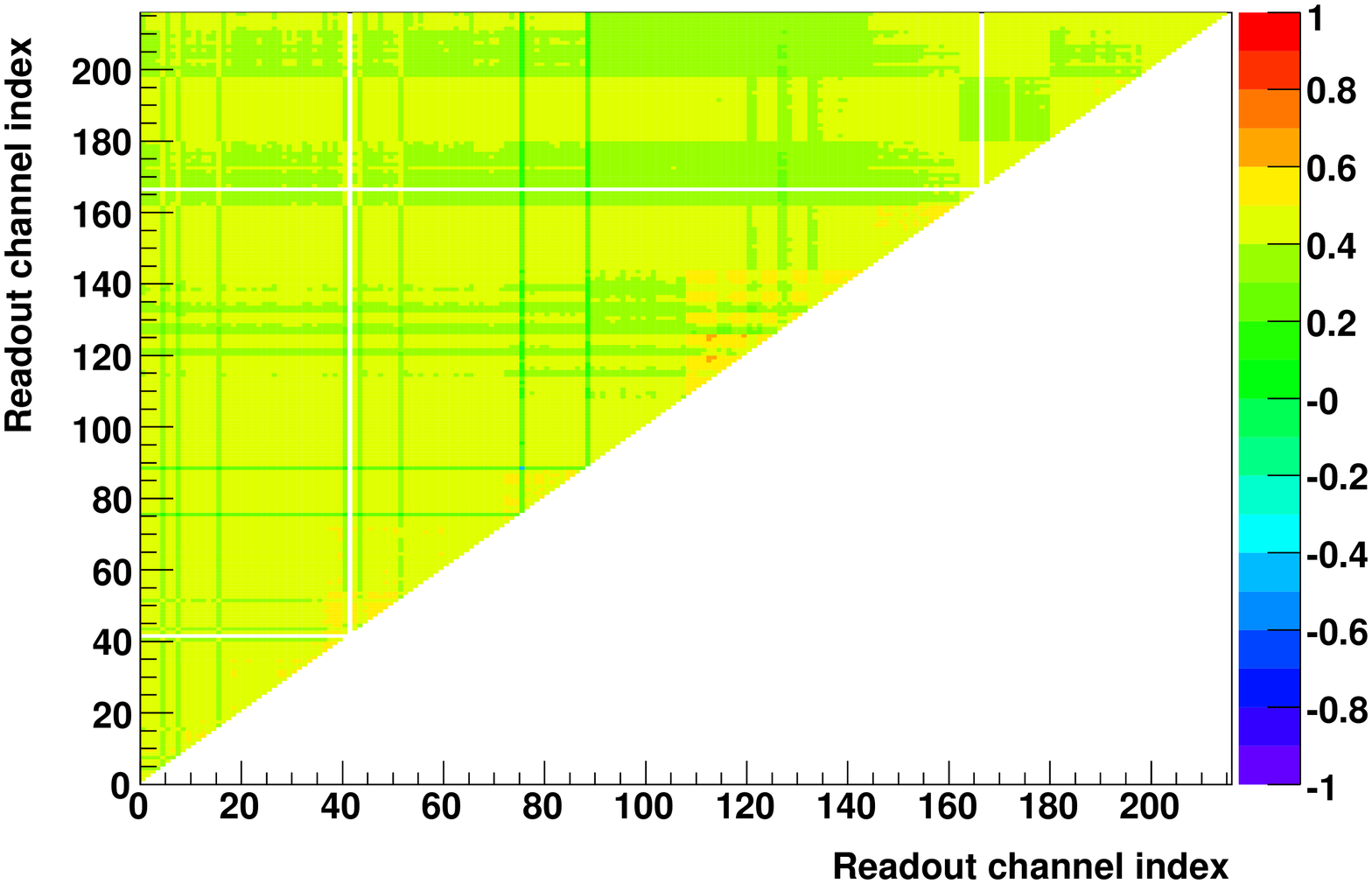}}
\end{minipage}
\hfill
\begin{minipage}{0.5\columnwidth}
\centerline{\includegraphics[angle=-90,width=0.98\columnwidth]{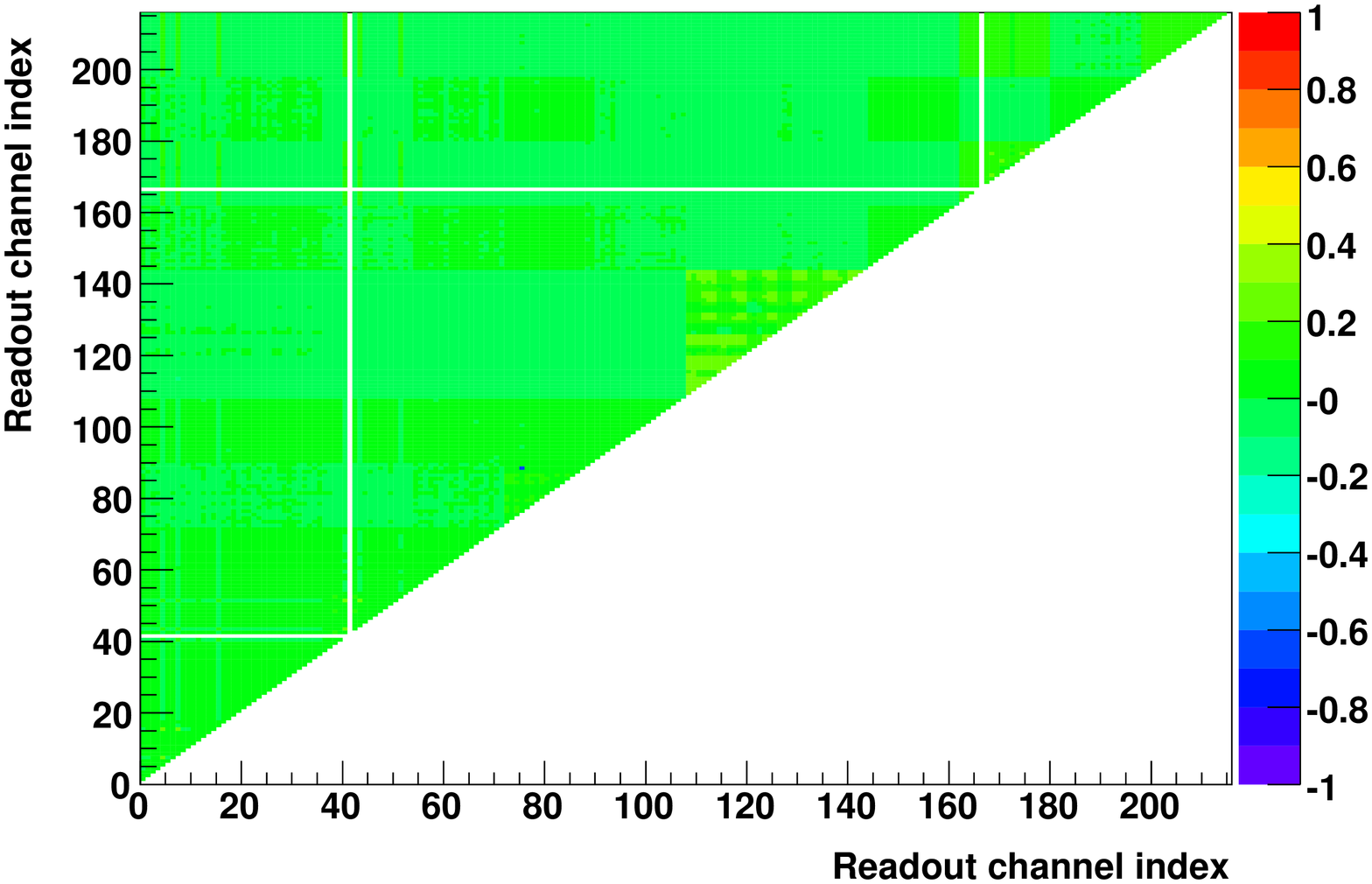}}
\end{minipage}
\caption{\sl Noise correlation coefficient (colour scale) between all pairs of channels for the VFE PCB situated in the third layer, in a 45\,$\mathrm{GeV}$ electron run taken at CERN in August 2006, after the simple pedestal subtraction. Left: before correcting the pedestal shifts, the mean correlation is $0.4207~\pm~0.0004$ with a standard deviation of $0.0551~\pm~0.0002$. Right: after corrections, the mean correlation is $-0.0017 \pm 0.0003$ with a standard deviation of $0.0519 \pm 0.0002$.}
\label{fig:cohNoise}
\end{figure}

In a first iteration, pedestal and noise values are obtained by calculating the mean and standard deviation of the ADC count distribution per channel in the 500 pedestal events taken at the start of each sequence. Averages calculated from a set of pedestal events are then simply subtracted from the ADC values recorded in the following set of beam events. The shift in the pedestals will affect the apparent noise, as the reference value is fixed by the previous set of pedestal events: the noise will appear to be artificially large, and 100\,\% correlated within the PCB if the shift is common to all channels. The left-hand plot of Figure~\ref{fig:cohNoise} shows the correlation coefficients (averaged over one run) between the apparent noise on the 216 channels of a particular VFE PCB, calculated after this simple pedestal subtraction. Channels are ordered by wafers, so 36 consecutive channels belong to the same wafer in the readout channel index displayed. The apparent noise is highly correlated between all channels, indicative of a common shift in the pedestal values.
%the mean correlation is $0.4207~\pm~0.0004$ with a standard deviation of $0.0551~\pm~0.0002$.

It is therefore important to correct the pedestals on an event-by-event basis, by detecting and correcting shifts compared with the previous estimate. Hence, a second iteration on the pedestal value is performed. A sample of channels is defined on each VFE PCB by selecting modules without signal hits. For this purpose, it is assumed that a signal was recorded if the difference between the minimum and the maximum of the pedestal-subtracted ADC values of the 36 channels in a module exceeds 3000\,ADC~counts. 
For this sample, a first estimate of the mean pedestal value is taken to be the average of the minimum channel value per chip, plus five times the noise. The root-mean-square (RMS) deviation from this mean is calculated using only the channels having ADC values smaller than the mean. This is to avoid being biased by the signal.
For each VFE PCB, a single offset is then estimated by iterating on the mean value (and hence the channels taken into account in the calculation of the RMS) until the RMS is compatible with the mean noise of the PCB, with a maximum of 20 iterations. The right-hand plot of Figure~\ref{fig:cohNoise} shows the noise correlation coefficients after applying this procedure. The mean correlation is reduced to $-0.0017 \pm 0.0003$, indicating that the procedure has successfully corrected for the shifts.

\subsubsection{Correlated noise and signal-induced pedestal shift}
\label{sips}

\begin{figure}[h!]
\begin{minipage}{0.5\columnwidth}
\centerline{\includegraphics[angle=-90,width=0.98\columnwidth]{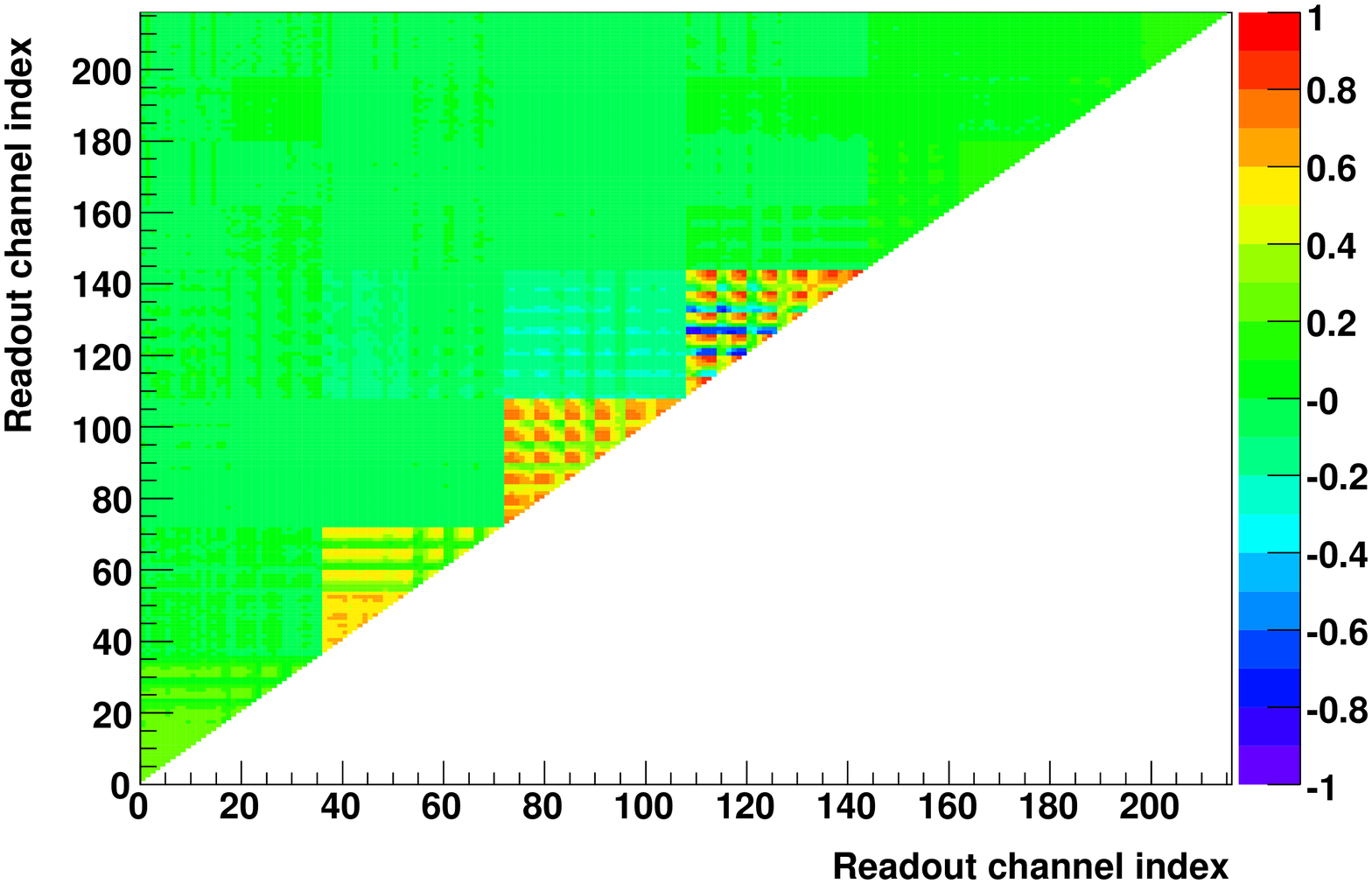}}
\end{minipage}
\hfill
\begin{minipage}{0.5\columnwidth}
\centerline{\includegraphics[angle=-90,width=0.98\columnwidth]{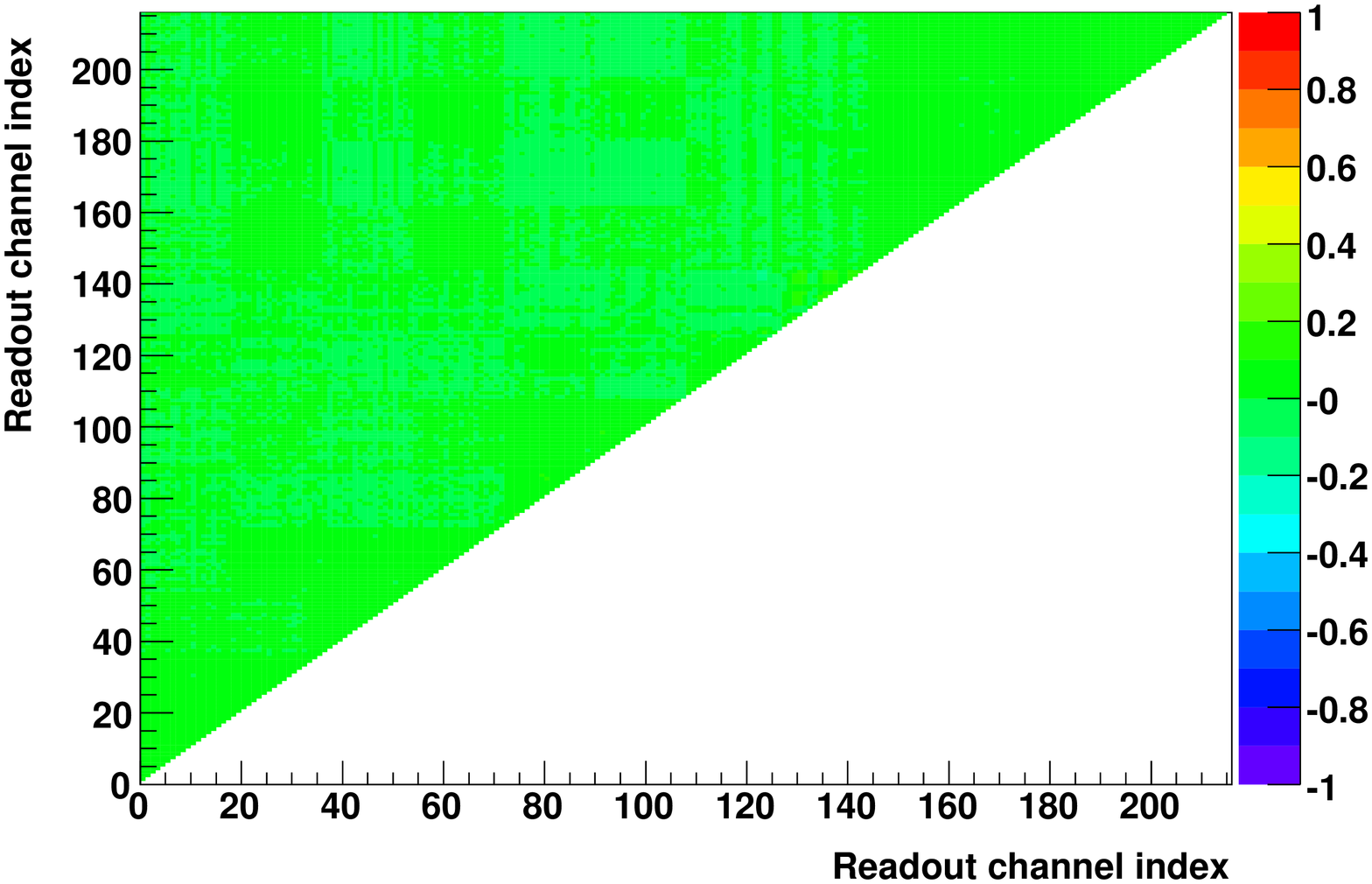}}
\end{minipage}
\caption{\sl Noise correlation coefficient (colour scale) between two channels for the VFE PCB of the tenth layer, in a 45\,$\mathrm{GeV}$ electron run taken at CERN in August 2006. Left: before SIPS corrections, the mean correlation is $0.019 \pm 0.001$ with a standard deviation of $0.156 \pm 0.001$. Right: after SIPS corrections, the mean correlation is $0.0058 \pm 0.0001$ with a standard deviation of $0.0194 \pm 0.0001$.}
\label{fig:crosstalk}
\end{figure}

Some modules recording a high signal are subject to a different effect, namely a signal-induced pedestal shift (SIPS), resulting in apparent high noise correlations for all 36 channels on a module. This can be seen in Figure~\ref{fig:crosstalk}, left, for the VFE PCB situated in the tenth layer, for the same run as in Figure~\ref{fig:cohNoise}, after applying the corrections described in Section~\ref{pedsubtr}.

However, the SIPS effects are not systematic, in that they do not affect all PCBs, and, for those affected, the effect is not always present. What is inducing this phenomenon is not yet understood, but a possible explanation could be that there is an intermittent bad contact between the 
\begin{wrapfigure}{l}{0.5\columnwidth}%[h!]
\centerline{\includegraphics[width=0.45\columnwidth]{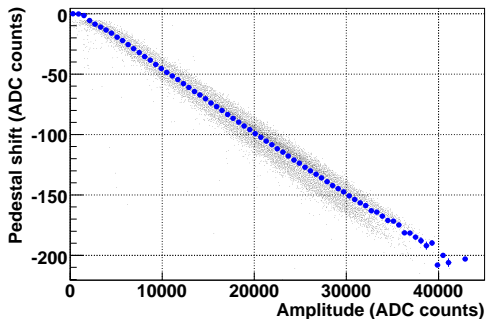}}
\caption{\sl Pedestal shift in the top row, middle module of the 11$^{th}$ layer as a function of the signal amplitude summed over all 36 channels, in a 45\,$\mathrm{GeV}$ electron run taken at CERN in August 2006. The average per bin in X is displayed in blue circles.}
\label{fig:xtalkintensity}
\end{wrapfigure}
aluminum foil and the module. A common serial resistance to all pads of a few Ohms would produce the same effect. A strong correlation between the negative pedestal shift and the amplitude of the signal is observed, as shown in Figure~\ref{fig:xtalkintensity}.

The SIPS are corrected on an event-by-event basis. Inspired by the corrections described in Section~\ref{pedsubtr}, averages are considered per module, rather than per PCB, discarding hits with signal, and iterating on the mean, standard deviation, and channels taken into account in the calculations. The result after corrections can be seen in Figure~\ref{fig:crosstalk}, right. The corrections perform well, the average correlation over the whole PCB being reduced by a factor of three, and the standard deviation of the correlation coefficients by one order of magnitude.

\subsection{Uniformity and stability of the residual pedestals and noise}
\label{pednoiseUnif}
%\label{defs}

For more systematic studies of uniformity and stability in time and temperature, the residual pedestal and noise will from now on be defined by the mean $R_G$ and standard deviation $N_G$ of the Gaussian fit of the noise component of the spectrum in beam induced events (see first peak in Figure~\ref{fig:cosmics}), after pedestal subtraction and all corrections as described above. Noise and residual pedestal values are verified to be stable within a run. The full statistics of each run is hence used to perform the fit, and the run dependence studied.

%There is however another way of extracting pedestal and noise values, by fitting the noise component of the spectrum in beam induced events with a Gaussian (see first peak in Figure~\ref{fig:cosmics}). This second method is used, after the pedestal subtraction and correction procedure, for more systematic studies of the noise correlations, uniformity and stability in time and temperature. 
The uncertainties on the residual pedestal $R_G$ and noise $N_G$ in the following sections arise from the uncertainties from the fit on each parameter. The noise peak is fitted iteratively 
% using the mean value and the standard deviation, 
in the range [$R_G-5\,N_G;R_G+1\,N_G$], to avoid being biased by the signal. By varying the fit range, an additional systematic uncertainty of $\pm 0.3 \% \times N_G$ is found and added in quadrature to the uncertainty on the noise.

%\subsubsection{Uniformity and stability of the residual pedestals}
%\label{pedUnif}
\subsubsection{Uniformity and stability of the residual pedestals}
\label{pedUnif}

\begin{figure}[h!]
\centerline{\includegraphics[width=0.8\columnwidth]{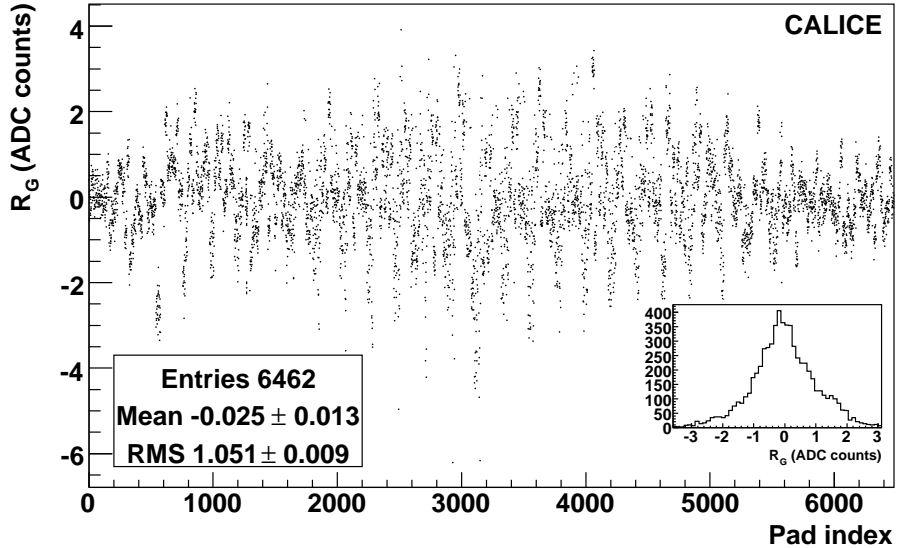}}
%\end{minipage}
\caption{\sl Uniformity of the residual pedestals $R_G$ as a function of the pad index (see text), for a 45\,$\mathrm{GeV}$ electron run taken at CERN in August 2006. The inset histogram displays the projection on the y-axis.}
\label{fig:pedperPix}
\end{figure}

Figure~\ref{fig:pedperPix} shows the residual pedestals for a particular run taken with electrons, as a function of the pad index. The pad index is defined as $6 \times 36 \times K + 36 \times (2\times W_C + W_R - 1) + (6\times P_C + P_R)$, where $K$ is the layer index, $W_R$ ($W_C$) is the module row (column) index, and $P_R$ ($P_C$) is the pad row (column) index. Results are not shown for the nine channels (out of 6471 active channels, see Section~\ref{calibresults}) where the Gaussian fit to the noise peak did not converge. The average residual pedestal over all channels is $-0.03 \pm 0.01$\,ADC~counts for this particular run, with a standard deviation of $1.05 \pm 0.01$\,ADC~counts. A larger scattering can be seen around the shower maximum, indicating a possible remaining influence of the SIPS described in Section~\ref{sips}. A periodic structure may be noted, whose period is 216 channels, i.e. PCB-related, and reflects the differences in length of the electronics connection from each pad to its readout chip.
%, and in Figure~\ref{fig:pedperChip} as a function of the time in seconds since the beginning of the run, for 18 channels of a randomly chosen module.

\begin{figure}[h!]
\begin{minipage}[l]{0.5\columnwidth}
\centerline{\includegraphics[width=0.98\columnwidth]{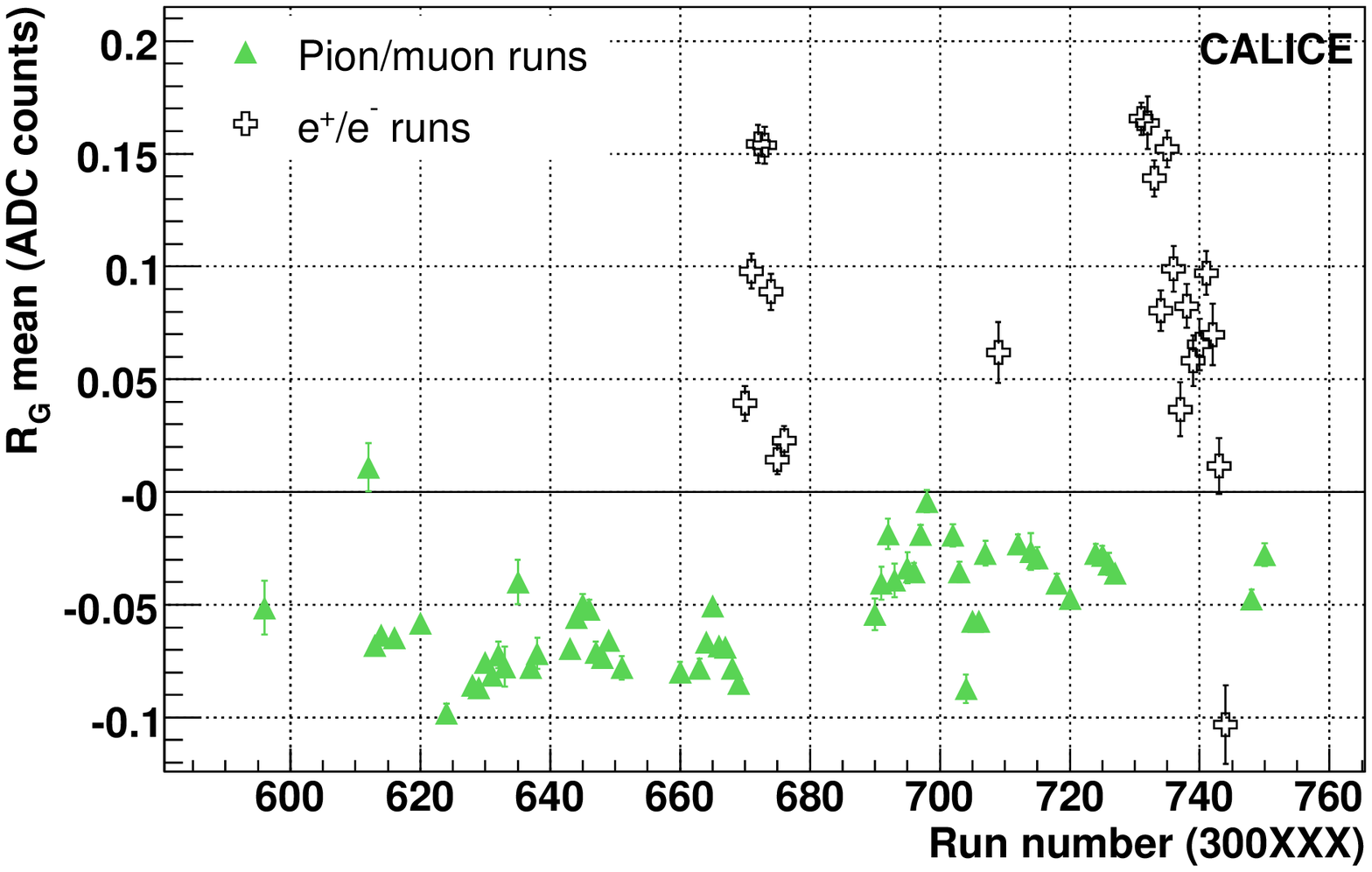}}
\end{minipage}
\hfill
\begin{minipage}[r]{0.5\columnwidth}
\centerline{\includegraphics[width=0.98\columnwidth]{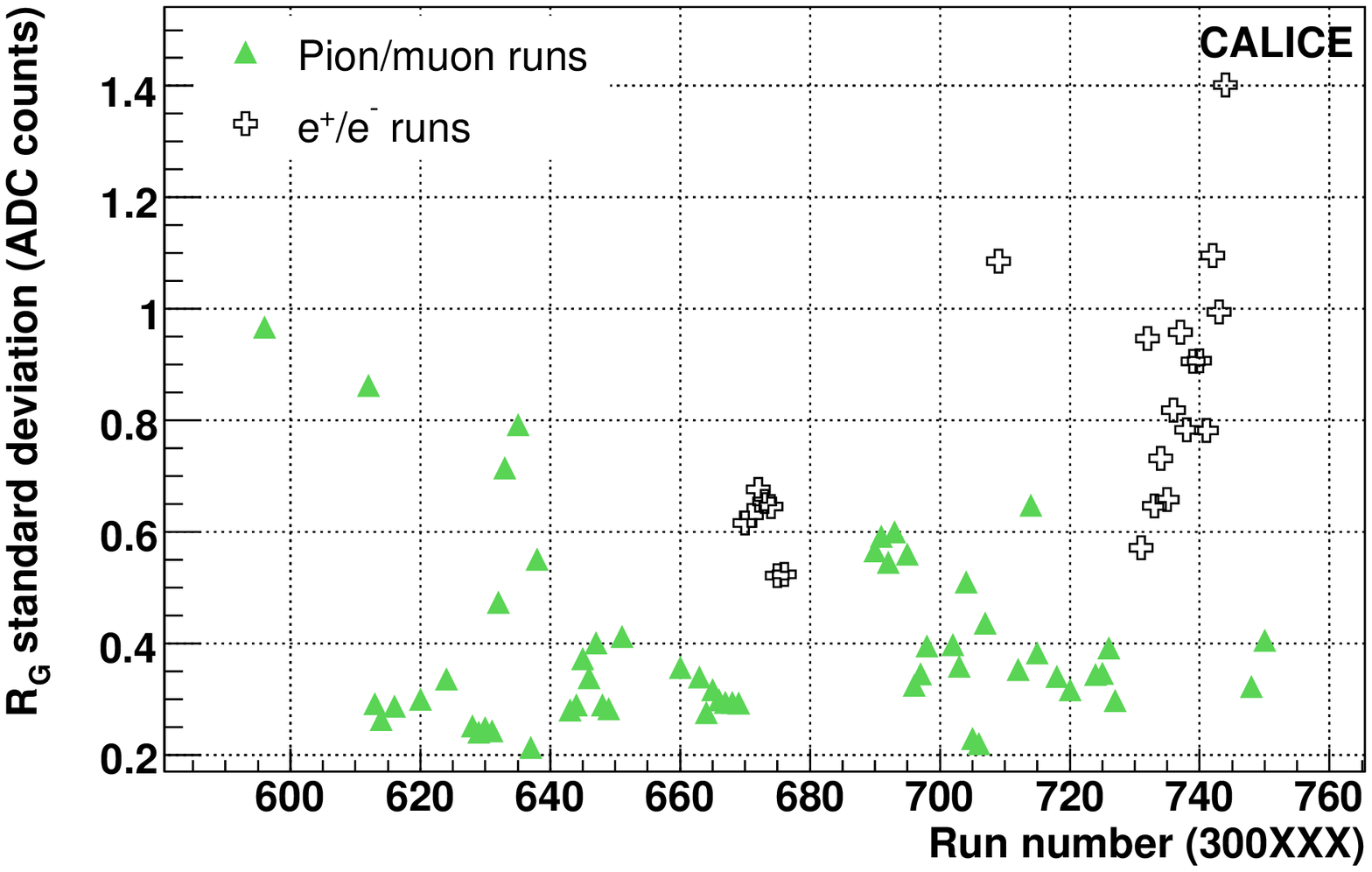}}
\end{minipage}
\caption{\sl  Time dependence of the average over all channels of the residual pedestal $R_G$: mean value (left) and standard deviation (right) as a function of the run number, for runs taken at CERN in October 2006.}
\label{fig:meanOct}
\end{figure}

Figure~\ref{fig:meanOct} shows the mean and standard deviation of the residual pedestals averaged over all channels, for other runs taken in October 2006, as a function of the run number. Very similar results are obtained for the August runs.
Pion and muon runs have systematically less spread between channels than electron runs, and a mean value closer to zero, indicating again a possible remaining influence of the SIPS on the pedestals.

%\begin{figure}[h!]
%\begin{minipage}[l]{0.5\columnwidth}
%\centerline{\includegraphics[width=0.98\columnwidth]{./sections/figures/meanPedvsRun_aug.eps}}
%\end{minipage}
%\hfill
%\begin{minipage}[r]{0.5\columnwidth}
%\centerline{\includegraphics[width=0.98\columnwidth]{./sections/figures/sigmaPedvsRun_aug.eps}}
%\end{minipage}
%\caption{\sl Time dependence of the average over all channels of the residual pedestal $R_G$: mean value (left) and standard deviation (right) as a function of the run number, for runs taken at CERN in August 2006.}
%\label{fig:meanAug}
%\end{figure}

%For the electron runs recorded in October 2006, the mean value for the offset is $0.080 \pm 0.013$\,ADC~counts with a spread run-to-run of $0.062 \pm 0.009$\,ADC~counts, and a standard deviation is $0.80\pm 0.05$\,ADC~counts with a spread run-to-run of $0.22 \pm 0.03$\,ADC~counts.
On average over all channels, the pedestals in electron runs are subtracted with a remaining positive offset of $0.080 \pm 0.013$\,ADC~counts ($0.20 \pm 0.02$\,\% of a MIP). The residual standard deviation channel-to-channel is however $0.80\pm 0.05$\,ADC~counts ($1.9 \pm 0.1$\,\% of a MIP), which justifies investigating the channel dependence in the following.

\begin{figure}[h!]
\begin{minipage}[l]{0.5\columnwidth}
\centerline{\includegraphics[width=0.98\columnwidth]{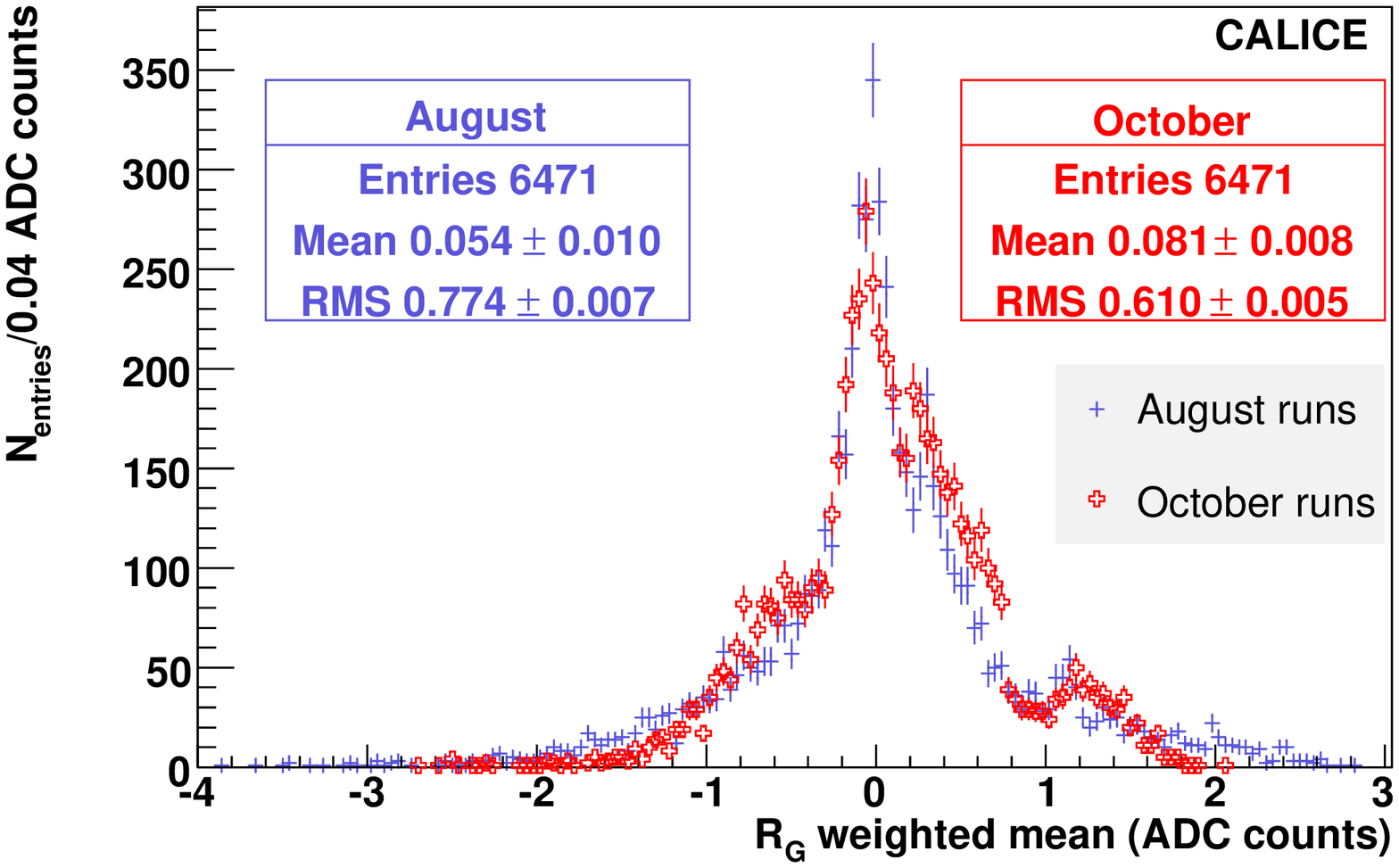}}
\end{minipage}
\hfill
\begin{minipage}[r]{0.5\columnwidth}
\centerline{\includegraphics[width=0.98\columnwidth]{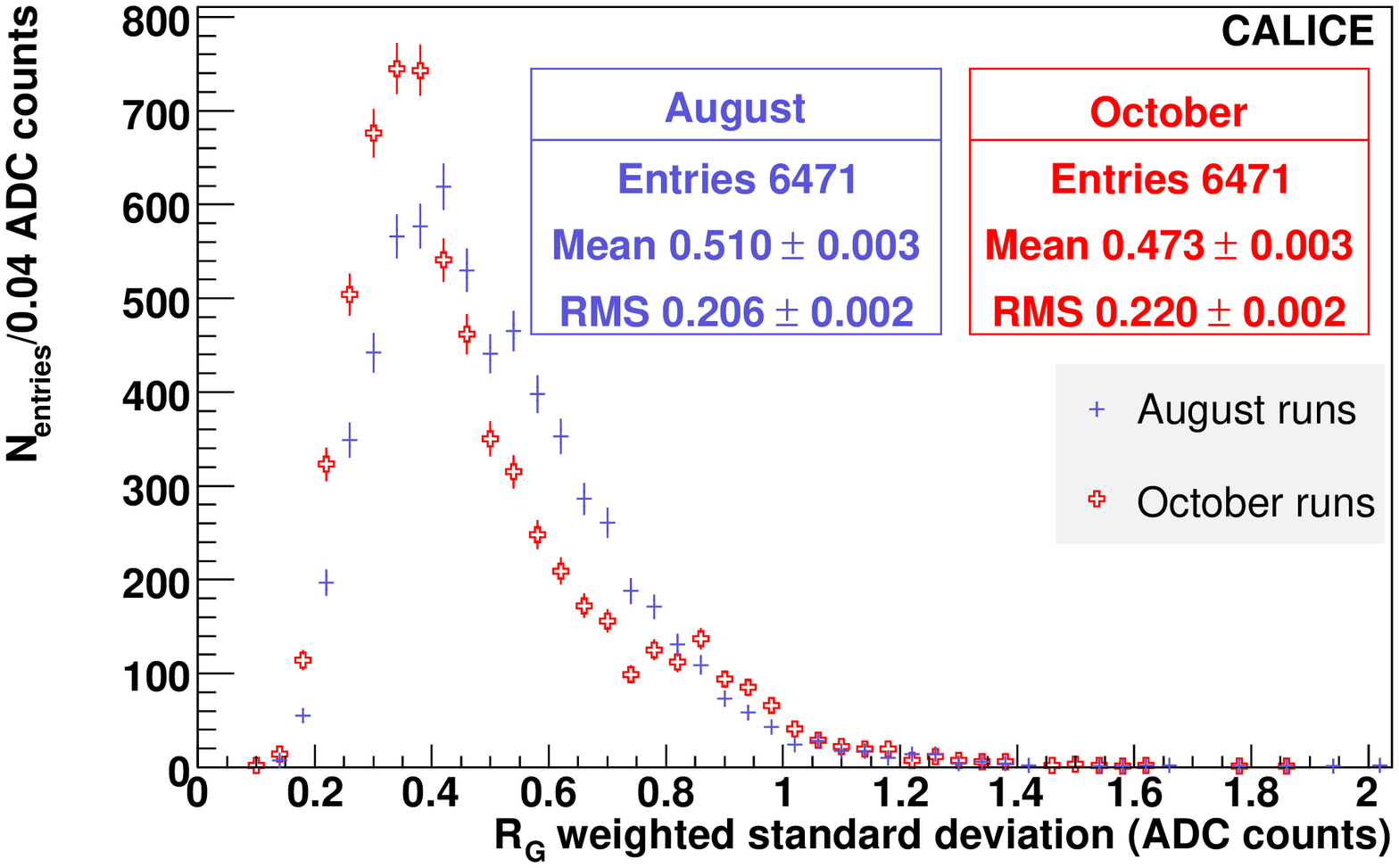}}
\end{minipage}
\caption{\sl Weighted averages over all runs of the residual pedestal $R_G$ per channel: weighted mean (left) and weighted standard deviation (right) for all channels, for electron runs taken at CERN in 2006.}
\label{fig:weightedMean}
\end{figure}

The residual pedestal per channel and per run is found to not have any clear dependence on time or temperature, with no observable day/night variations, nor any dependence on the beam energy of the run.
In order to study more systematically the time dependence for each channel, the mean and standard deviation over all runs weighted by the error per run are considered per channel, and displayed in Figure~\ref{fig:weightedMean} for all channels, for CERN electron runs.
The mean pedestal offset is smaller than $0.17 \pm 0.02$\,\% of a MIP with a standard deviation channel-to-channel of $1.67 \pm 0.02$\,\% of a MIP. The residual offset run-to-run is of the order of $1.1 \pm 0.1$\,\% of a MIP, with a standard deviation channel-to-channel of $0.48 \pm 0.01$\,\% of a MIP.

The effect of a systematic pedestal offset on the energy scale is additive, and so the mean offset scales with the number of hits recorded. To give an estimate of the size of the effect, a 1\,$\mathrm{GeV}$ (45\,$\mathrm{GeV}$) electron will produce around 250\,MIPs (6000\,MIPs). If these are distributed over 100 (400) hits, the impact of the remaining pedestal offset will be around 0.2\,MIP (0.8\,MIP), i.e. $\simeq$1\,MeV ($\simeq$3\,MeV), which can be neglected.

%measured at $0.054 \pm 0.010$ ($0.081 \pm 0.008$)\,ADC~counts with a standard deviation channel-to-channel of $0.774 \pm 0.007$ ($0.610 \pm 0.005$)\,ADC~counts, with a spread between runs of $0.511 \pm 0.003$ ($0.473 \pm 0.003$)\,ADC~counts with a standard deviation channel-to-channel of $0.207 \pm 0.002$ ($0.220 \pm 0.002$)\,ADC~counts, for the August (October) 2006 electron runs. The offset is hence smaller than 0.2\,\% of a MIP, with channel-to-channel variations of the order of 1.7\,\% of a MIP, and run-to-run variations of $1.1 \pm 0.1$\,\% of a MIP.

%\begin{figure}[h!]
%\centerline{\includegraphics[width=0.95\columnwidth]{./sections/figures/weightedMean_oct.eps }}
%\centerline{\includegraphics[width=0.95\columnwidth]{./sections/figures/weightedMeanSpread_oct.eps}}
%\caption{\sl Weighted average (top) and standard deviation (bottom) over all runs of the pedestal per channel for all channels, for electron runs taken at CERN in October 2006.}
%\label{fig:weightedMeanOct}
%\end{figure}

%\subsection{Noise - 2.5p - AMM+MR}
%\label{noise}

%\subsubsection{Coherent noise and crosstalk}
%\label{noise}

\subsubsection{Uniformity and stability of the noise}
\label{noiseUnif}

\begin{figure}[h!]
\centerline{\includegraphics[width=0.9\columnwidth]{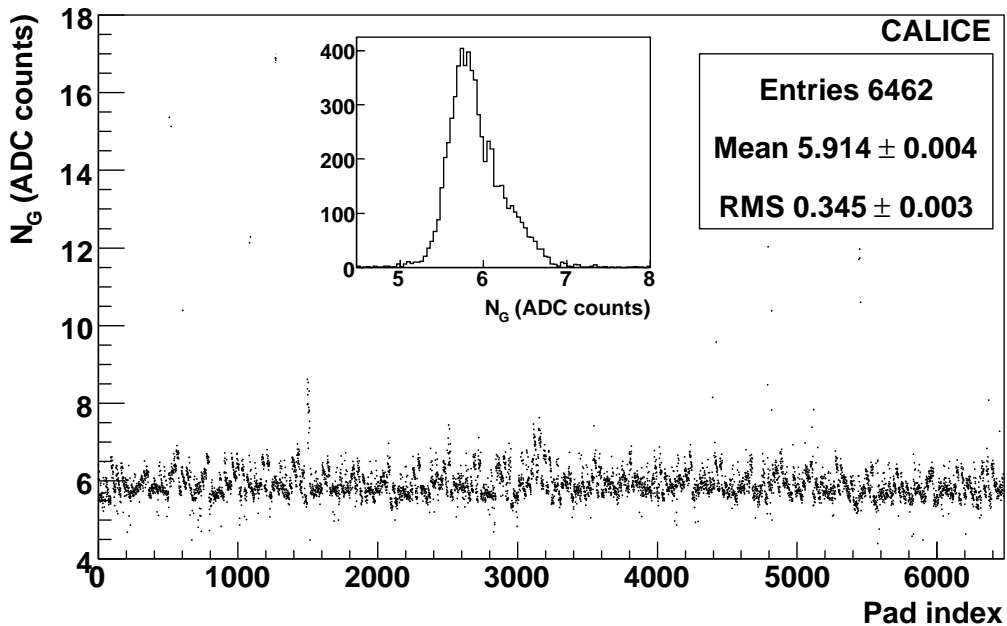}}
%\end{minipage}
\caption{\sl Uniformity of the noise $N_G$ as a function of the pad index (see text) for a 45\,$\mathrm{GeV}$ electron run. The inset histogram displays the projection on the y-axis.}
\label{fig:rmsperPix}
\end{figure}

%\begin{wrapfigure}{l}{0.5\columnwidth}%[h!]
%\begin{minipage}[l]{0.48\columnwidth}
Figure~\ref{fig:rmsperPix} shows the noise per channel as a function of the same pad index as used in Figure~\ref{fig:pedperPix}, and for the same electron run. Some channels have an intrinsically high noise, often in (anti)correlation with a neighbouring channel. On average over all channels, the noise with this method is found to be $5.914 \pm 0.004$\,ADC~counts with a standard deviation of $0.345 \pm 0.003$\,ADC~counts, for this particular run.

%\begin{figure}[h!]
%\begin{minipage}[l]{0.5\columnwidth}
%\centerline{\includegraphics[width=0.98\columnwidth]{./sections/figures/meanNoisevsRun_aug.eps}}
%\end{minipage}
%\hfill
%\begin{minipage}[l]{0.5\columnwidth}
%\centerline{\includegraphics[width=0.98\columnwidth]{./sections/figures/sigmaNoisevsRun_aug.eps}}
%\end{minipage}
%\caption{\sl  Time dependence of the average over all channels of the noise $N_G$: mean value (left) and standard deviation (right) as a function of the run number, for runs taken at CERN in August 2006.}
%\label{fig:rmsAug}
%\end{figure}

\begin{figure}[h!]
\begin{minipage}[l]{0.5\columnwidth}
\centerline{\includegraphics[width=0.98\columnwidth]{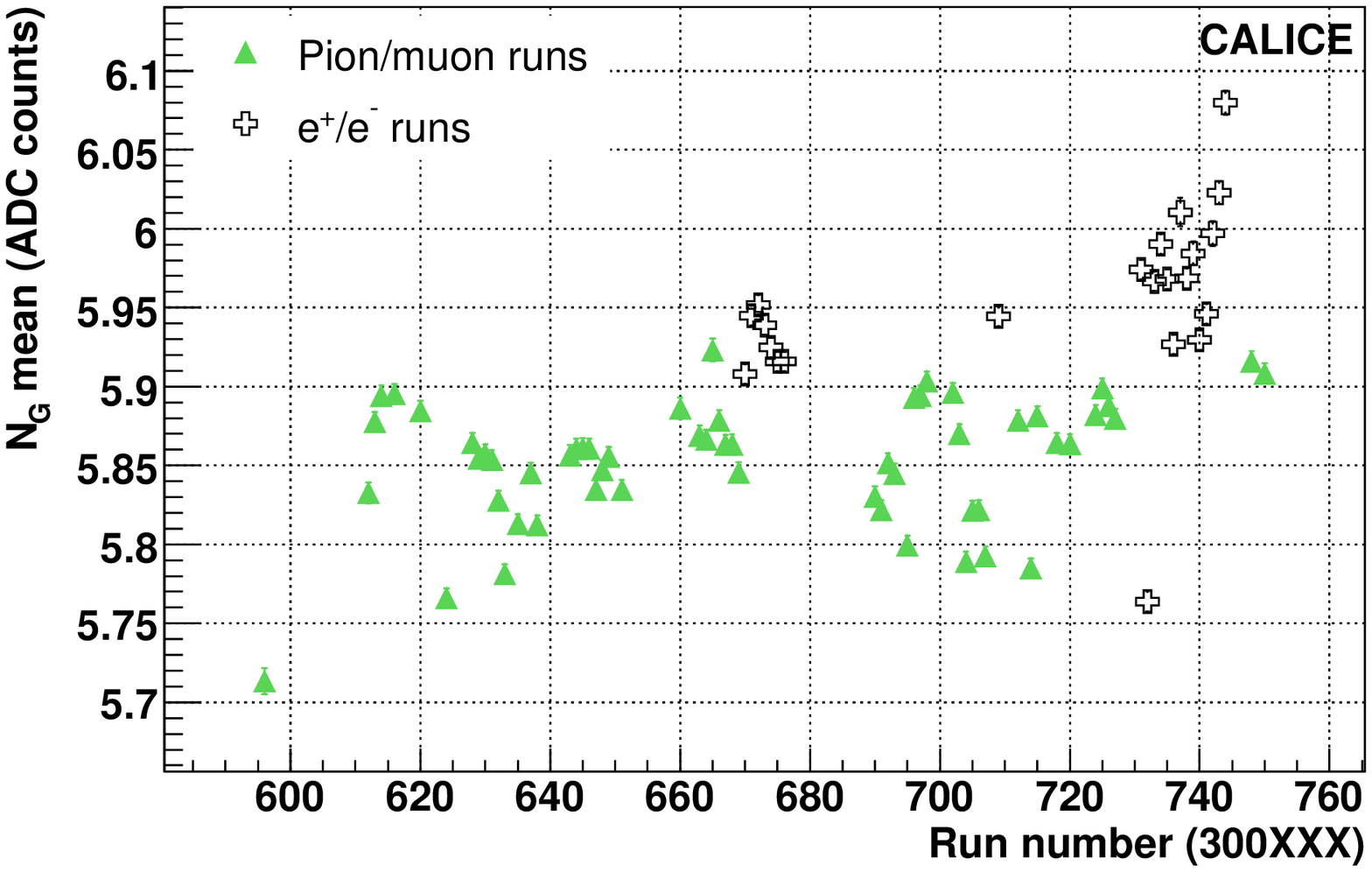}}
\end{minipage}
\hfill
\begin{minipage}[l]{0.5\columnwidth}
\centerline{\includegraphics[width=0.98\columnwidth]{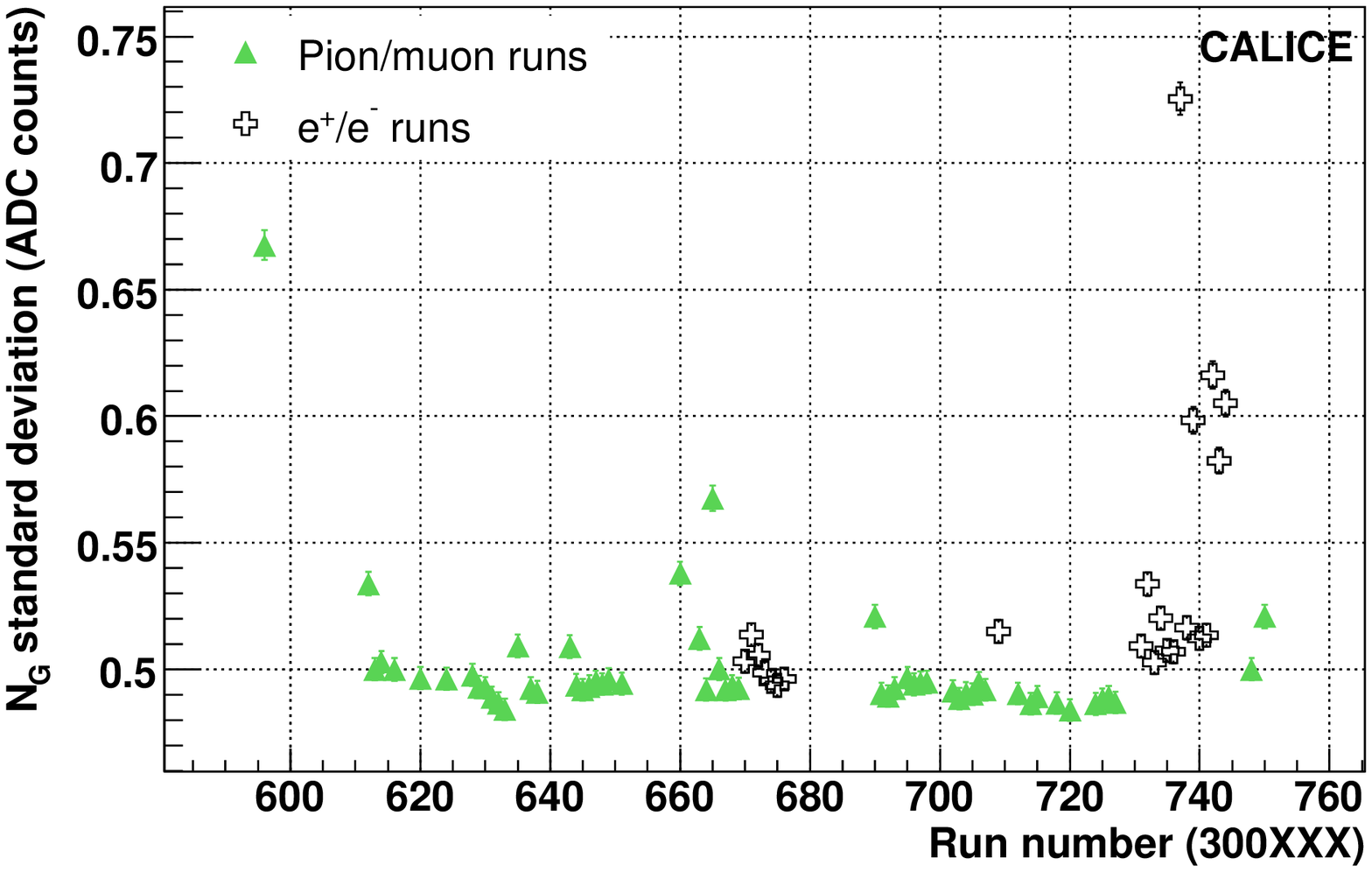}}
\end{minipage}
\caption{\sl Time dependence of the average over all channels of the noise $N_G$: mean value (left) and standard deviation (right) as a function of the run number, for runs taken at CERN in October 2006.}
\label{fig:rmsOct}
\end{figure}

Figure~\ref{fig:rmsOct} shows the mean and standard deviation of the noise averaged over all channels, for other runs taken in October 2006, as a function of the run number. Very similar results are obtained for the August runs.
%Discarding pion and muon runs from the results presented in Figure~\ref{fig:rmsOct}, the 21 remaining electron runs give a mean value for the noise at $5.962 \pm 0.009$\,ADC~counts with a spread run-to-run of $0.041 \pm 0.006$\,ADC~counts, and a standard deviation of $0.54\pm 0.01$\,ADC~counts with a spread run-to-run of $0.057 \pm 0.009$\,ADC~counts. Very similar results are found for the August runs.
On average over all channels, the noise is $5.962 \pm 0.009$\,ADC~counts ($12.9 \pm 0.1$\,\% of a MIP), with a run-to-run dependence smaller than 1\,\% of the mean noise, and a channel-to-channel dependence of about 9\,\% of the mean noise, which justifies considering the channel dependence in the following.

\begin{figure}[h!]
%\begin{minipage}[l]{0.45\columnwidth}
%\centerline{\includegraphics[width=1.09\columnwidth]{./sections/figures/noisePerRunnum_Aug.eps}}
%\hfill
%\centerline{\includegraphics[width=0.95\columnwidth]{./sections/figures/rmsPerTime_Oct.eps}}
%\caption{\sl Time dependence of the mean value of the noise contribution in signal events, for five individual channels representatively chosen across the detector, for runs taken at CERN in August (top) and in October (bottom) 2006.}
%\caption{\sl Time dependence of the noise per channel for three individual channels representatively chosen across the detector : noise $N_G$ as a function of the run number, for runs taken at CERN in August 2006.}
%\label{fig:rmsperChan}
%\end{minipage}
%\hfill
\begin{minipage}[l]{0.5\columnwidth}
\centerline{\includegraphics[width=0.98\columnwidth]{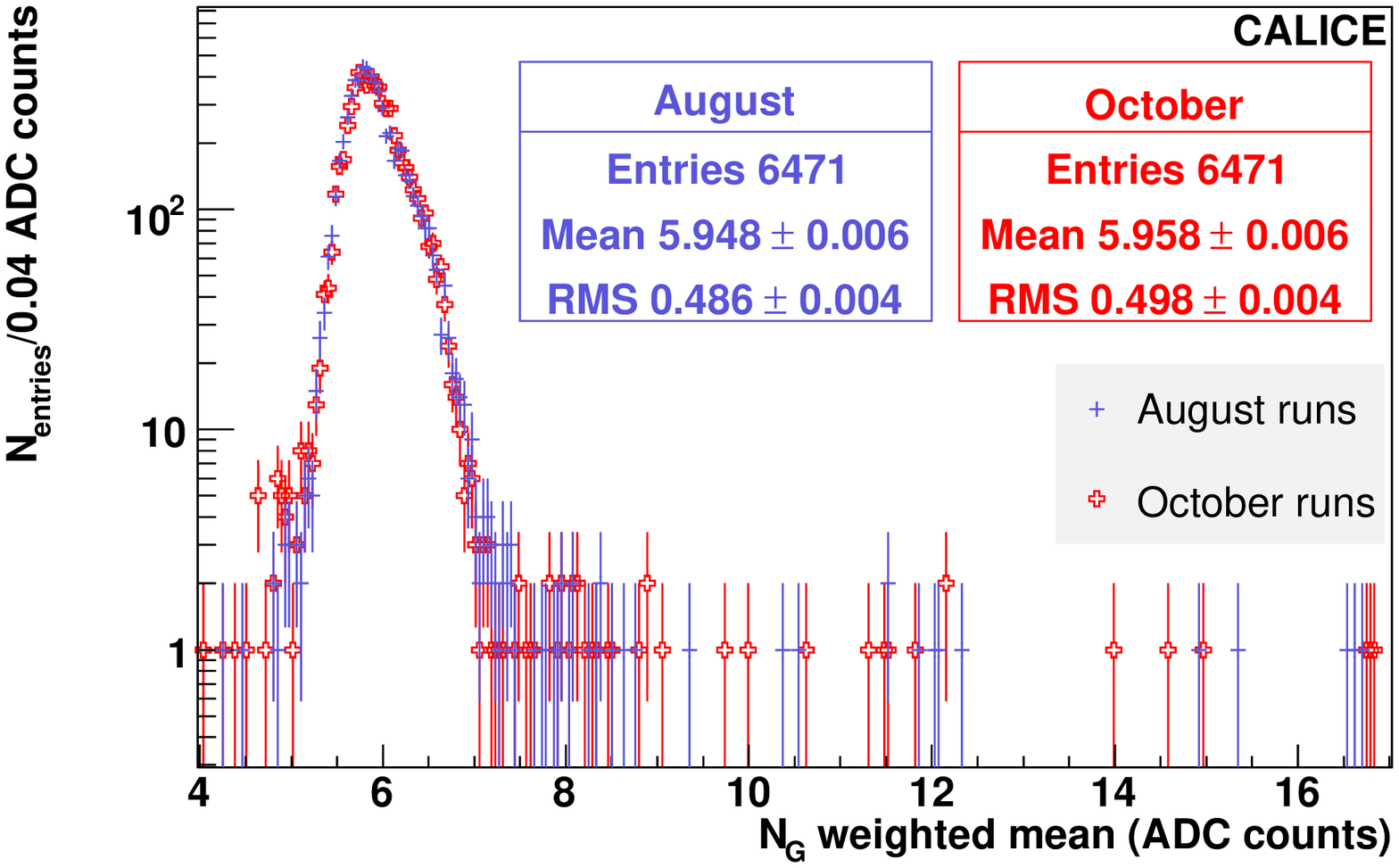}}
%\label{fig:weightedRMS}
\end{minipage}
%\end{figure}
%\begin{figure}[h!]
%\begin{minipage}[l]{0.5\columnwidth}
%\centerline{\includegraphics[width=0.98\columnwidth]{./sections/figures/weightedRmsSpread.eps}}
%\centerline{\includegraphics[width=0.95\columnwidth]{./sections/figures/weightedRms_oct.eps }}
%\end{minipage}
\hfill
\begin{minipage}[r]{0.5\columnwidth}
%\begin{figure}[h!]
\centerline{\includegraphics[width=0.98\columnwidth]{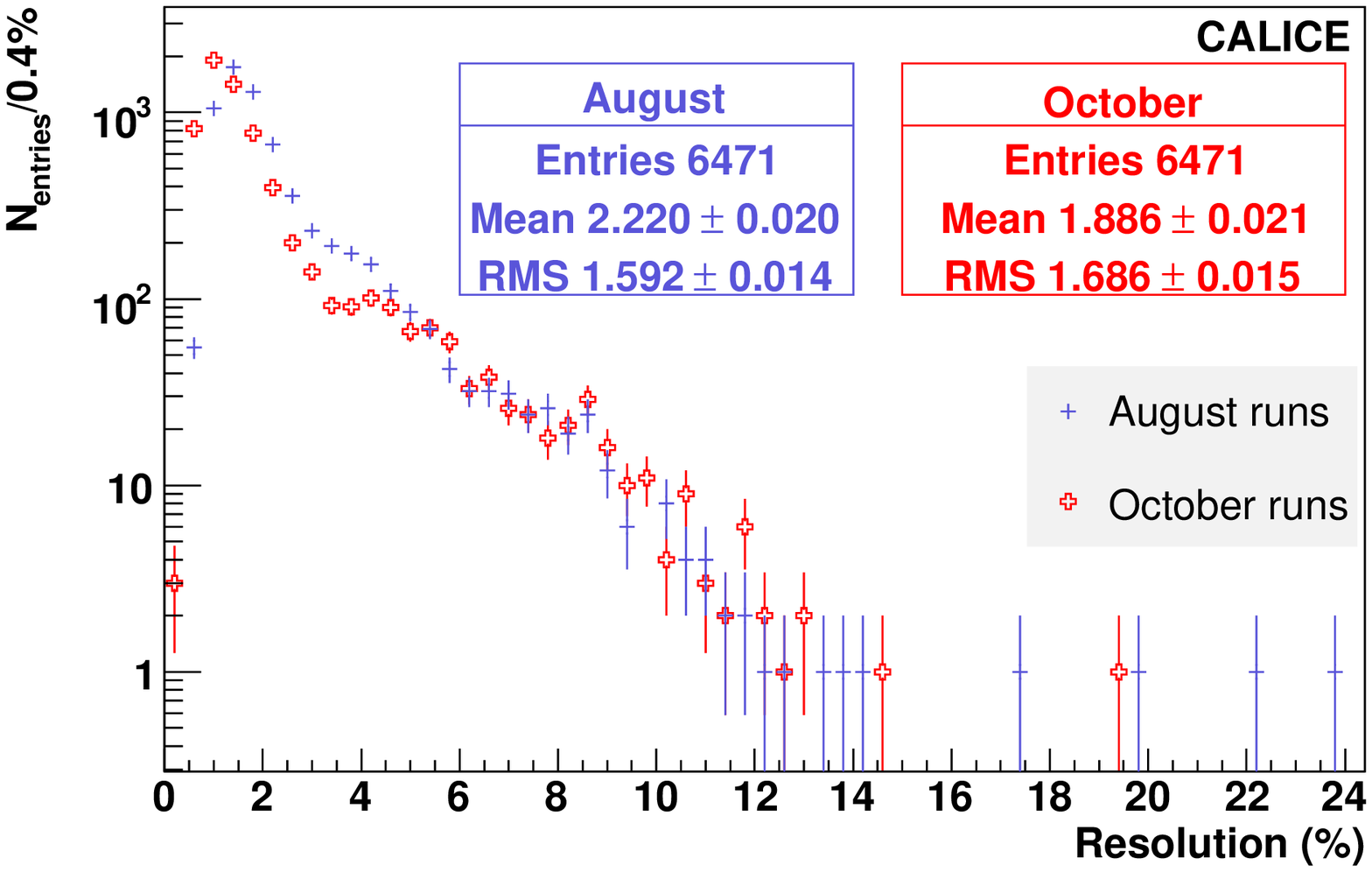}}
%\centerline{\includegraphics[width=0.95\columnwidth]{./sections/figures/weightedRmsSpreadRel_oct.eps}}
\end{minipage}
\caption{\sl Weighted mean over all runs (left), and resolution, defined as the ratio of the weighted standard deviation to the weighted mean (right), of the noise $N_G$ per channel for all channels, for electron runs taken at CERN in 2006.}
%\caption{\sl Weighted standard deviation (left) over all runs, and ratio of the standard deviation by the mean (right), of the noise $N_G$ per channel for all channels, for electron runs taken at CERN in 2006.}
\label{fig:weightedRMSSpreadRel}
\end{figure}

%The noise per channel is shown in Figure~\ref{fig:rmsperChan}, for the same three individual channels as in Figure~\ref{fig:pedTimeDep}, as a function of the run number, for several runs taken at CERN in August 2006. As for the residual pedestals, no clear correlations are seen as a function of the temperature (day/night variations) or the beam energy.%, as seen in Figure~\ref{fig:rmsvsE} for the latter.
As for the residual pedestal, no clear correlations are seen as a function of time, temperature (day/night variations), or beam energy.
%\begin{figure}[h!]
%\centerline{\includegraphics[width=0.95\columnwidth]{./sections/figures/rmsPerE_Aug.eps}}
%\caption{\sl Energy dependence of the noise per channel in signal events, for five individual channels representatively chosen across the detector, for runs taken at CERN in August 2006.}
%\label{fig:rmsvsE}
%\end{figure}
In order to study more systematically the time dependence for each channel, the mean noise value over all runs weighted by the error per run is considered per channel, and displayed for all channels in Figure~\ref{fig:weightedRMSSpreadRel}, left, for CERN electron runs.
The mean noise is $12.9 \pm 0.1$\,\% of a MIP, with a spread channel-to-channel of $1.1 \pm 0.1$\,\% of a MIP. The standard deviation in time is smaller than 0.3\,\% of a MIP on average, with a spread channel-to-channel of less than 0.2\,\% of a MIP. 
%On average, the noise is measured at $5.948 \pm 0.006$  ($5.958 \pm 0.006$)\,ADC~counts with a standard deviation channel-to-channel of $0.486 \pm 0.004$ ($0.498 \pm 0.004$)\,ADC~counts, and a spread between runs of $0.132 \pm 0.001$ ($0.112 \pm 0.001$)\,ADC~counts with a standard deviation channel-to-channel of $0.0945 \pm 0.0008$ ($0.1026 \pm 0.0009$)\,ADC~counts, for the August (October) 2006 electron runs. 
The standard deviation per channel normalised by the corresponding noise per channel is represented in Figure~\ref{fig:weightedRMSSpreadRel}, right. On average over both periods, the relative spread run-to-run is $2.00 \pm 0.03$\,\% with a spread between channels of $1.60 \pm 0.01$\,\%. With about 20\,\% of the channels having run-to-run variations above 3\,\%, a run-by-run approach to measure the noise is required.

%To quantify the impact of the noise on the energy scale, the channel-to-channel and run-to-run variations of the pedestal described in Section~\ref{pedUnif} must be added in quadrature to the mean noise. The resulting noise is found at $6.07 \pm 0.05$\,ADC~counts ($13.2 \pm 0.1$\,\% of a MIP).
The impact of the noise on the energy scales with the square root of the number of hits: for 100 (400) hits in the case of a 1\,$\mathrm{GeV}$ (45\,$\mathrm{GeV}$) incident electron, the noise contributes less than 0.5\,\%/$\sqrt{E}$ to the total energy resolution.

\subsubsection{Conclusion}

%Averages in time and over channels of the residual pedestals and noise are summarised in Table~\ref{tab:pednoise}.
Pedestals are subtracted with a remaining offset smaller than 0.2\,\% of a MIP, but with channel-to-channel variations of $1.7 \pm 0.1$\,\% of a MIP, and run-to-run variations of $1.1 \pm 0.4$\,\% of a MIP. The impact on the energy scale depends on the number $N$ of hits recorded, $0.002 \times N$ in MIP units, which can be neglected. 

 The noise per channel is $12.9 \pm 0.1$\,\% of a MIP on average for all channels, with variations in time averaged over all channels of $2.00 \pm 0.03$\,\%. About 20\,\% (6\,\%) of the channels have time variations greater than 3\,\% (5\,\%), requiring a run-by-run approach to extract the noise measurements. The impact on the energy resolution also depends on the number $N$ of hits recorded, $0.13 \times \sqrt{N}$ in MIP units. Since $N \simeq 100$ (400) for 1\,$\mathrm{GeV}$ (45\,$\mathrm{GeV}$), this is therefore negligible.

%\begin{table}[h!]
%\caption{\sl Summary of the averages in time and over all channels of the residual pedestals and noise in October 2006 CERN electron data: mean and standard deviation ($\sigma$) for both periods. The averages of $R_G$ and $N_G$ over all channels have corresponding averages in time, and the weighted averages of $R_G$ and $N_G$ in time have corresponding averages over all channels.}
%\vspace{0.2cm}
%\label{tab:pednoise}
%\begin{center}
%\begin{tabular}{|c|c|c||c|c|} 
%\cline{2-5}
%\multicolumn{1}{c|}{}  & August & October & August & October \\
%\cline{2-5}
%\multicolumn{1}{c|}{} & \multicolumn{4}{c|}{Residual pedestals (ADC counts)}\\
%\cline{2-5}
%\multicolumn{1}{c|}{}  & $R_G$ $mean_{channels}$ & $R_G$ $\sigma_{channels}$ & $N_G$ $mean_{channels}$ & $N_G$ $\sigma_{channels}$ \\
%\cline{2-5}
%\hline
%$mean_{time}$ & 0.080 $\pm$ 0.013 & 0.80 $\pm$ 0.05 & 5.962 $\pm$ 0.009 & 0.54 $\pm$ 0.01 \\
%%\hline
%$\sigma_{time}$ & 0.062 $\pm$ 0.009 & 0.22 $\pm$ 0.03 & 0.041 $\pm$ 0.006 & 0.057 $\pm$ 0.009 \\
%\hline
%\multicolumn{1}{c|}{}  & $mean_{channels}$ & $\sigma_{channels}$  & $mean_{channels}$ & $\sigma_{channels}$ \\
%\hline
%$R_G$ $[mean_{time}]_{weighted}$ &  0.081 $\pm$ 0.008 & 0.610 $\pm$ 0.005 &  5.958 $\pm$ 0.006 & 0.498 $\pm$ 0.004 \\
%$R_G$ $[\sigma_{time}]_{weighted}$ &  0.473 $\pm$ 0.003 & 0.220 $\pm$ 0.002 & 0.113 $\pm$ 0.001 & 0.103 $\pm$ 0.001 \\
%\hline
%\end{tabular}
%\end{center}
%\end{table}

\subsection{Crosstalk around the module edge}
\label{squareEvents}

\begin{figure}[h!]%{l}{0.5\columnwidth}
\begin{minipage}[l]{0.45\columnwidth}
%\centerline{\includegraphics[width=1.09\columnwidth]{./sections/figures/SQWexp.eps}}
%\caption{\sl Example of square pattern in a module: energy deposits (colour scale) in layer 6 as a function of the position of the hits in the layer, for a particular event (5261) of a 30\,$\mathrm{GeV}$ electron run (300378) taken at CERN in 2006.}
%\label{fig:SQWexp}
\centerline{\includegraphics[width=1.09\columnwidth]{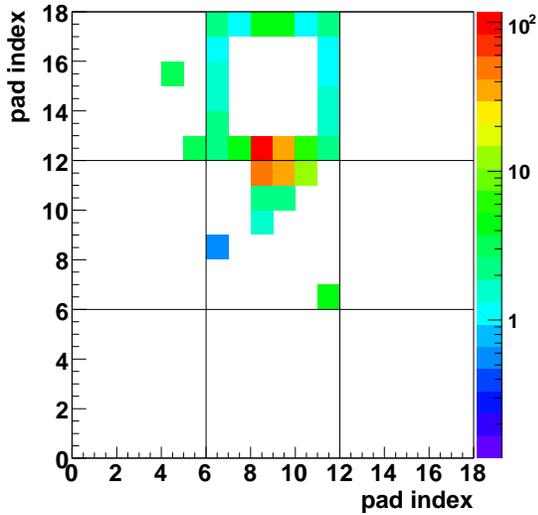}}
\caption{\sl Example of a square event, taken at CERN with a 45\,$\mathrm{GeV}$ electron beam at normal incidence, in the 12$^{th}$ layer. The colour scale represents the energy deposited per pad, in units of MIP.}
\label{fig:SQE}
%\end{figure}
%\begin{figure}[h!]
\end{minipage}
\hfill
\begin{minipage}[l]{0.45\columnwidth}
\centerline{\includegraphics[width=1.09\columnwidth]{./sections/figures/GRmodel.eps3}}
\caption{\sl Simple electrical model of capacitive coupling of pads to the guard rings.}
\label{fig:GRmodel}
\end{minipage}
\end{figure}

In some events (as shown in the event display in Figure~\ref{fig:SQE}), when a large quantity of energy is deposited in the proximity of the guard rings, a signal of almost constant amplitude appears on all peripheral pads, except for the four pads of the corners where this amplitude is twice as large. These ``square events'' are due to capacitive coupling which exists between the guard rings and the peripheral pads. From the measurements, about 1\% of the charge deposited in the guard ring is propagated into each border pixel (and twice as much into the corner pixels), which is compatible with the simulation of a simple electrical model shown in Figure~\ref{fig:GRmodel}.

Following the idea of a capacitive coupling between guard rings and bordering pads, a straightforward option is to cut the guard rings into small segments. This layout technique should result in the coupling capacitance being in series, thus dividing the coupling effects. Following preliminary simulation studies, the segmented topology design option for the guard rings is shown to lower the crosstalk effects by a factor ranging from 3 to 30 according to the actual distances of the elements of the design. The simulation work is ongoing and other design options are being evaluated. Measurements on real modules will complete this work.

The criterion for identifying square events in the beam test data is based on two characteristics of the border hits involved in square patterns. Firstly, a border hit is considered as isolated if its neighbours are only other border hits. Secondly, two border hits are linked if there is a path between them along the border without any gap. In a module, a square pattern is found each time the number of border hits which are isolated and linked is greater than nine, established with the help of Monte Carlo data. The Monte Carlo does not include the simulation of the crosstalk.

\begin{figure}[h!]
%\begin{wrapfigure}{r}{0.6\columnwidth}
\centerline{\includegraphics[width=0.6\columnwidth]{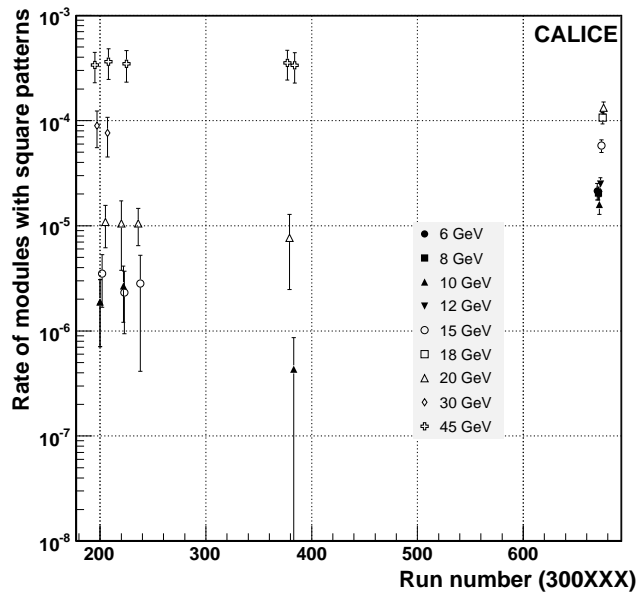}}
\caption{Square pattern rate per module for a particular algorithm (see text), for runs taken at CERN in August and October 2006. Energies from 6 to 45\,$\mathrm{GeV}$ are displayed.}
\label{fig:RawRateAugOct06}
\end{figure}

The probability for a module to display a square pattern according to the above criterion is shown in Figure~\ref{fig:RawRateAugOct06} for both August and October periods of data taking at CERN. The error bars displayed in Figure~\ref{fig:RawRateAugOct06} contain the systematic uncertainty associated with varying the selection cut between eight and ten linked border hits, in quadrature with the statistical uncertainty. It is seen that, as expected for a crosstalk effect with a detection threshold, the rate increases with the energy of the electron beam, but remains stable in time for a given energy and beam profile. For both periods, the beam was pointing towards a guard ring, but its geometrical spread was larger in August. This leads to a higher rate of square patterns in October, since the rate also increases when the particle becomes closer to the guard ring. Given this geometrical consideration, it is not reliable to quote a rate of square events independent of a particular beam setting.

%\subsection{Calibration}
\section{Calibration of the detector}
\label{calib}
%The determination of the pedestals and noise for the individual channels, as well as the calibration of the detector, i.e. the conversion of ADC counts to MIP equivalent energies, are described in the following sections.

Muons are used to calibrate the detector in terms of MIPs. After checking the stability in time of the noise component in muon events, the conversion factor from ADC~counts to MIP equivalent energies needs to be extracted per channel, using a large sample of muon data to reduce the statistical fluctuations. The uniformity and the stability in time of the calibration constants need to be checked, as well as the possible sources of systematic uncertainties that will influence any energy measurement. The method and results are detailed in the following sections.

\subsection{Extraction of the calibration constants per channel}
\label{calibintro}
\begin{figure}[h!]
\centerline{\includegraphics[width=0.7\columnwidth]{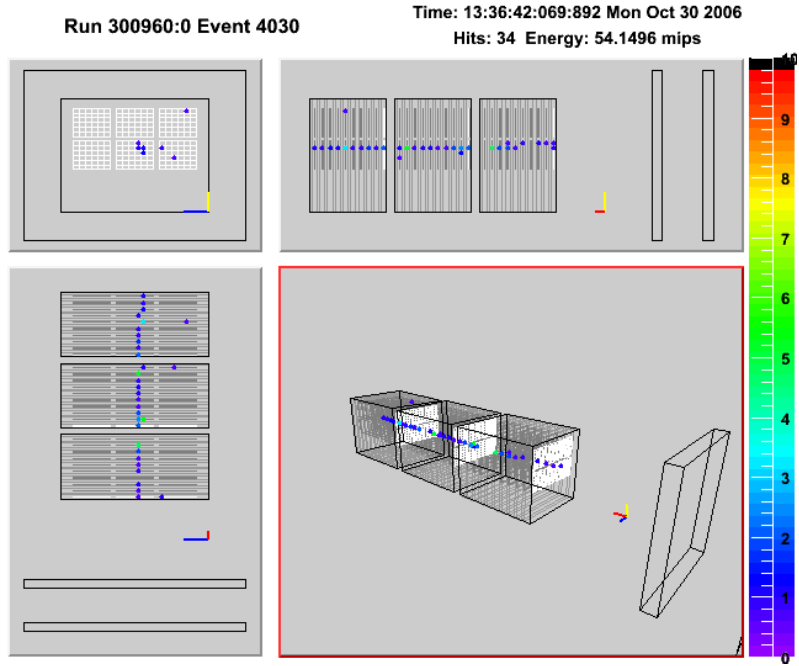}}
\caption{\sl Event display showing a muon crossing all 30 layers of the ECAL, in a run taken at CERN in October 2006. A threshold of 0.5\,MIP has been applied. The colour scale represents the energy deposited per pad, in units of MIP. Top-left: $X-Y$ projection, top-right: $Y-Z$ projection, bottom-left: $X-Z$ projection.}
\label{fig:muDisp}
\end{figure}

The calibration of the ECAL prototype was performed using 74 beam halo muon runs ($\simeq 250,000$ events each), recorded during October 2006. A further 18 runs were recorded in August 2006. Another experiment upstream helped to provide a wide spread of the muon beam over the whole surface of the prototype. Events were triggered with a 1 m$^2$ scintillator counter. An event display of a muon event is shown in Figure~\ref{fig:muDisp}.

%\subsubsection{Pedestal subtraction in muon events}
%\label{calibintro}
The first stage in the analysis of muon events is substract the pedestals from the raw data, as described in Section~\ref{pedestal}. 
The pedestal subtraction is performed with an accuracy of 0.1\,ADC~counts (0.2\,\% of a MIP) on average for all channels, taken as a systematic uncertainty for the calibration constants.

%\subsubsection{Fitting procedure}
%\label{calibmeth}

The calibration constants are calculated for each channel using the hit energy distribution in identified beam halo muon events. The event sample is selected by requiring the presence of a track which is consistent with being a MIP, based on two criteria. Firstly, the total number of reconstructed hits in the 30 layers of the ECAL must be between 15 and 40. Secondly, the distance between two hits in consecutive layers must be less than 2\,cm. 
As the residual pedestals are shown to be negligible compared to the estimated MIP signal (0.2\,\% of a MIP), they are not subtracted from the measured value of the MIP peak position. Furthermore, with a mean noise of $12.9 \pm 0.1$\,\% of a MIP, the noise tail is neglected and only the MIP contribution in the hit energy distribution is fitted.

\begin{figure}[h!]
%\begin{minipage}[l]{0.45\columnwidth}
\centerline{\includegraphics[width=0.6\columnwidth]{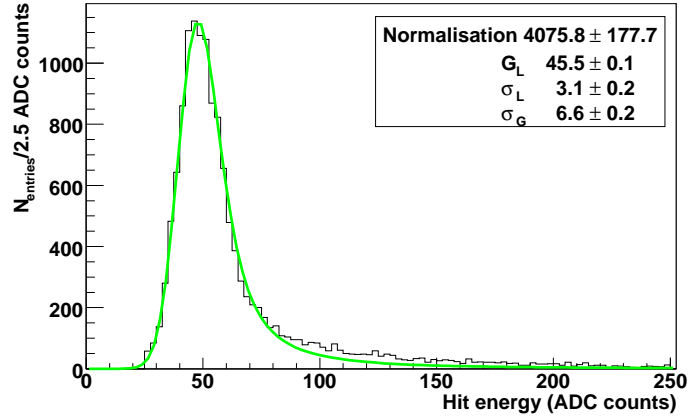}}
%\end{minipage}
%\hfill
%\begin{minipage}{3.8cm}
%\centerline{\includegraphics[width=0.98\columnwidth]{../Marcel/crap_fit.eps}}
%\end{minipage}
\caption{\sl A representative example of the energy distribution of hits in muon events for a particular channel. The fit function is a convolution of a Gaussian and a Landau. Normalisation, $G_{L}$ and $\sigma_{L}$ refers to the constant, most probable value and width of the Landau function, and $\sigma_{G}$ to the width of the Gaussian function.}
\label{fig:fit}
%\end{minipage}
%\hfill
%\begin{figure}[h!]
%\begin{minipage}[r]{0.45\columnwidth}
%\centerline{\includegraphics[width=1.09\columnwidth]{./sections/figures/FitErrorOnCalibConsts.eps}}
%\caption{\sl Error on the fitted value of $G_L$ for all channels, in October 2006 muon runs.}
%\label{fig:fitError}
%\end{minipage}
\end{figure}

For each channel, the distribution of hit energies (in ADC~counts) is made from the sample of selected muons. Since the muon spread is not totally uniform over the whole surface, the number of hits in the distributions varies from a few thousand in the border regions to up to 14,000 in the central part. The hit energy distribution is fitted to a convolution of a Landau distribution and a Gaussian. An example fit is shown in Figure~\ref{fig:fit}. The most probable value $G_L$ of the Landau function gives the calibration constant, while the standard deviation of the Gaussian $\sigma_{G}$ gives an estimate of the noise value for each channel. 
%The high statistics allow a small bump at around 100\,ADC~counts (approximately twice the estimated MIP value) to be seen, which can be explained by the emission of $\delta$-rays. 
%The high energy tail is badly reproduced by the fit, which can be explained by the emission of $\delta$-rays. The fitting range has thus been limited between 25 and 78.5\,ADC~counts.
The fitting range has been limited between 25 and 78.5\,ADC~counts.

The statistical uncertainty on the fitted value of $G_L$, calculated by TMinuit \cite{TMinuit}, is $0.243 \pm 0.001$\,ADC~counts on average for all channels, with a spread between channels of $0.063 \pm 0.001$ ADC~counts. In addition, as stated above, the pedestal subtraction introduces an additional uncertainty of $\pm 0.1$\,ADC~counts. If the entire ADC range is fitted rather than the range 25-78.5 ADC~counts, the difference is found to be $0.00 \pm 0.02$\,ADC~counts on average over all channels, with a spread between channels of $0.146 \pm 0.001$\,ADC~counts. The latter value is considered as a systematic uncertainty on the calibration constants $G_L$. The total statistical uncertainty on each calibration constant is hence about 0.24\,ADC~counts (0.5\,\% of a MIP), and the total systematic uncertainty 0.18\,ADC~counts (0.4\,\% of a MIP).
%computed by the sum in quadrature of these three sources of errors.

\subsection{Results}
\label{calibresults}

For 6403 out of the 6480 channels of the prototype, the MIP calibration is obtained using the above procedure. One entire module (36 channels) shows no signal above 25\,ADC~counts, as a result of the silicon not being fully depleted. The ratio between the mean MIP signals of this module and a randomly chosen neighbour is estimated to be 0.517, allowing a relative calibration of its channels. Nine channels have readout issues, and are declared dead. For the remaining 32 channels, the fit described above fails due to anomalously high levels of noise. 18 of these are recovered by fitting with the sum of a Gaussian and the previously used function. The other 14 are calibrated using neighbouring pads. The calibration constants are found to have slight variations chip-to-chip. The spread between chips is found to be $0.78 \pm 0.02$\,ADC~counts (1.7\,\% of a MIP) on average, hence this value is taken as systematic uncertainty for the 50 channels calibrated using neighbouring pads.

\subsubsection{Uniformity across the detector}

\begin{figure}[h!]
\centerline{\includegraphics[width=0.98\columnwidth]{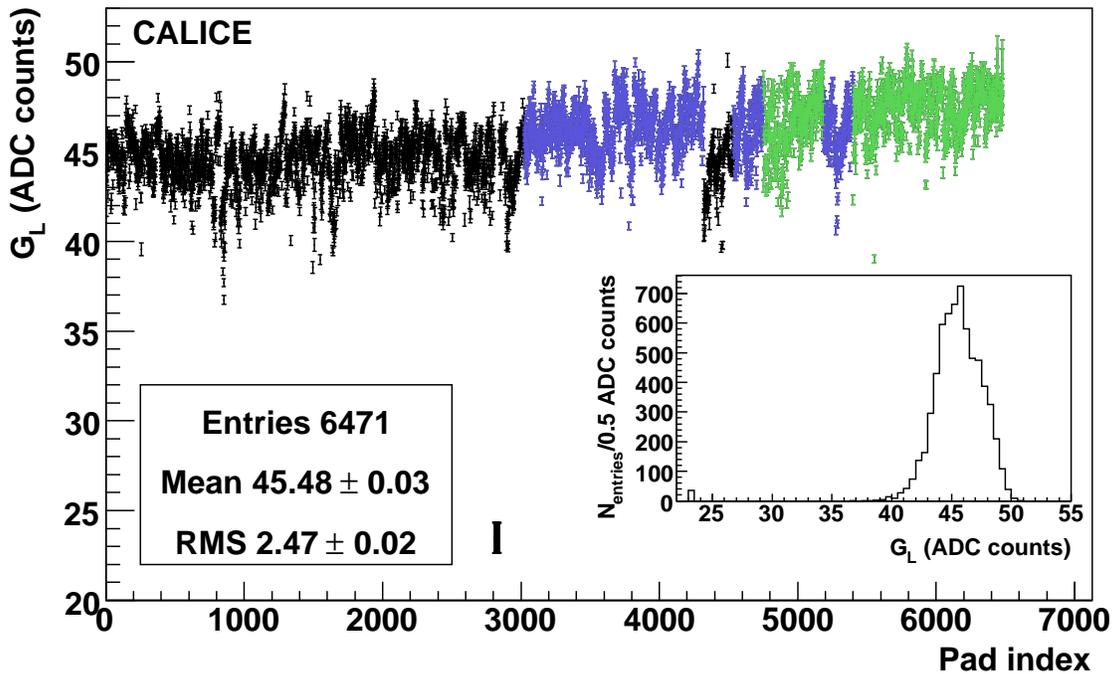}}
\caption{\sl Equivalent ADC value $G_L$ for the energy deposit of a MIP for all 6471 active channels of the ECAL prototype, obtained using the October 2006 muon events sample, as a function of the pad index. Points displayed in black, blue and green correspond to channels respectively from the first, second and third categories explained in the text. The group of 36 channels at 23.5\,ADC~counts corresponds to the non-fully depleted module. The inset histogram displays the projection on the y-axis.}
\label{fig:calib}
\end{figure}

The calibration constants for all channels are shown in Figure~\ref{fig:calib} as a function of the pad index. Three main categories of channels can be identified in Figure~\ref{fig:calib}, linked to the gluing date and the module origin. The first category (in black) at around 44\,ADC~counts contains layers 0 to 13 and layer 20 and corresponds to wafers produced by the Institute of Nuclear Physics, Moscow State University~\cite{Moscow}, glued at the end of 2004. The second category (in blue) at around 46\,ADC~counts contains layers 14 to 19, layer 21 and layer 24, and corresponds to wafers from the same manufacturer glued between October 2005 and May 2006.
The third category (in green) at around 47.5\,ADC~counts contains layers 22 and 23, and layers 25 to 29, and corresponds to wafers produced by the second manufacturer (ON Semiconductor Czech Republic~\cite{ONsemi}, with the Institute of Physics, Academy of Sciences of the Czech Republic, Prague), glued in 2006.

\subsubsection{Stability in time}
\label{calibtime}
To check the stability in time of the calibration constants, the values obtained with the above sample of beam halo muons recorded in October 2006 are compared with the test bench measurements, taken before the mounting of the slabs (see Section~\ref{slabCalib}). They are also compared with a sample of beam halo muons recorded in August 2006 in similar conditions.

\begin{figure}[h!]
\begin{minipage}{0.45\columnwidth}
\centerline{\includegraphics[width=1.09\columnwidth]{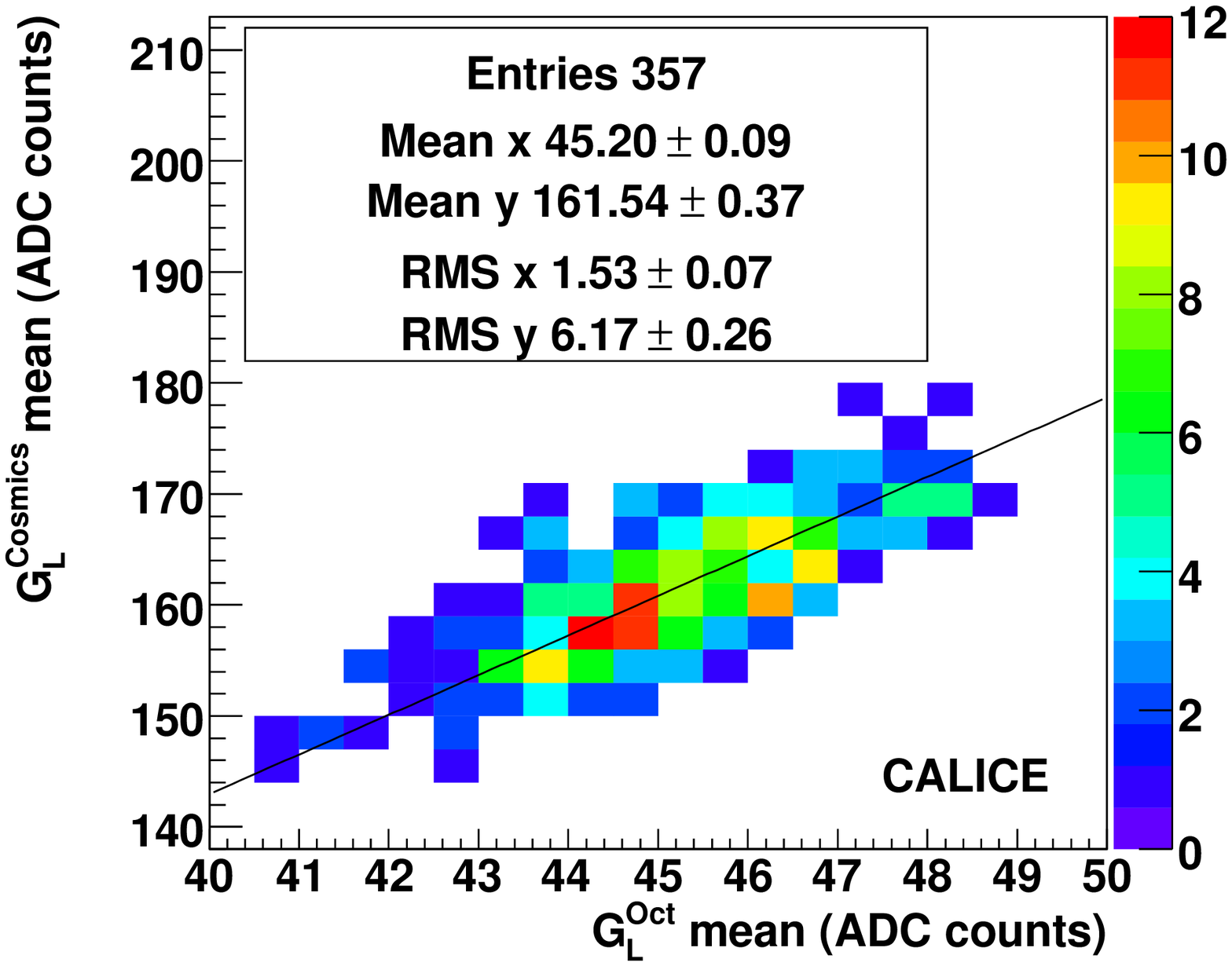}}
\caption{\sl Mean per chip of the calibration constants obtained with the cosmics test bench versus those obtained with October 2006 muon runs, discarding two outliers. The ADC~counts represented in both axis are obtained with different DAQ systems. The colour scale represents the number of entries per bin. The black line represents the equation $y=\frac{161.54}{45.20}\times x$. The values for the half-depleted module are out of scale for clarity, but are approximately aligned with the black line. One outlier chip at (43,260) is not shown for clarity.
%The non-fully depleted module is out of scale for clarity, but would appear at 87\,ADC~counts in the y-axis and 23.5 in the x-axis. 
}
\label{fig:benchvsoct}
\end{minipage}
\hfill
\begin{minipage}{0.45\columnwidth}
\centerline{\includegraphics[width=1.09\columnwidth]{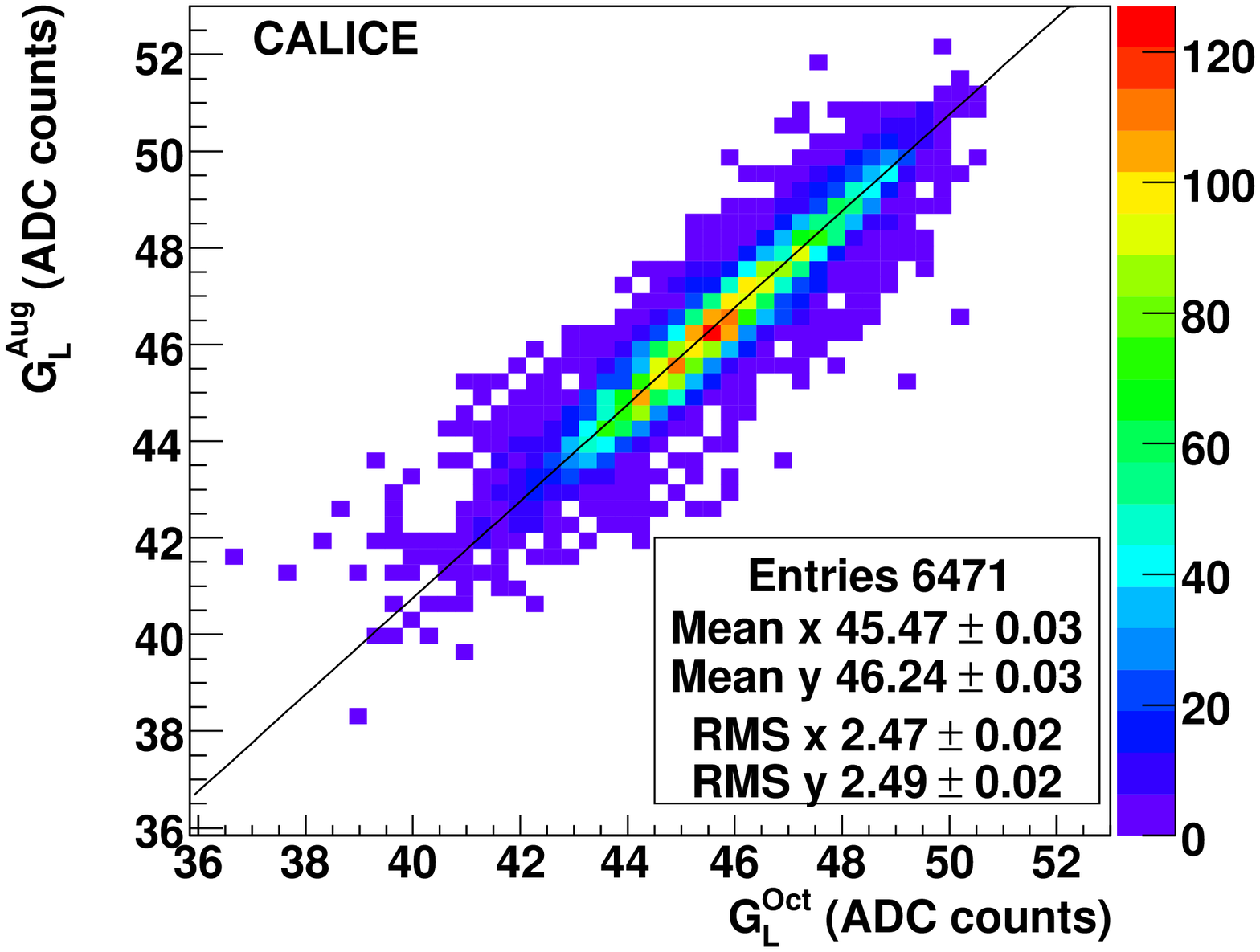}}
\caption{\sl Calibration constants obtained with August 2006 muon runs versus those obtained with October 2006 muon runs. The colour scale represents the number of entries per bin. The black line represents the perfect linearity for the two sets, taking into account a shift of 0.76\,ADC~counts ($y=x+0.76$). The values for the half-depleted module are not shown but are perfectly aligned with the black line.}
\label{fig:augvsoct}
\end{minipage}
\end{figure}

%The results from October are also compared with the results obtained with the test bench before mounting the slabs. 
Test bench measurements were made between 2004 and 2006 (see Section~\ref{slabCalib}), giving an average per chip (i.e. per 18 channels), with an uncertainty of 5\,\% (2.2\,ADC~counts on average), and with a different DAQ. These are plotted against the average per chip calculated for the October 2006 muon runs in Figure~\ref{fig:benchvsoct}. 
%\begin{figure}[h!]
%\begin{minipage}{0.5\columnwidth}
%\centerline{\includegraphics[width=0.98\columnwidth]{./sections/figures/CalibOctvsCosmics.eps}}
%\end{minipage}
%\hfill
%\begin{minipage}{0.5\columnwidth}
%\centerline{\includegraphics[width=0.98\columnwidth]{./sections/figures/CalibOctvsCosmicsZoom.eps}}
%\end{minipage}
%\caption{\sl Mean per chip of the calibration constants obtained with the cosmics test bench versus those obtained with October 2006 muon runs, for all chips (left) and discarding the three outliers (right). The ADC~counts represented in both axis are obtained with different DAQ systems. The colour scale represents the number of entries per bin. 
%The non-fully depleted module is out of scale for clarity, but would appear at 87\,ADC~counts in the y-axis and 23.5 in the x-axis. 
%The black line represents the equation $y=\frac{161.54}{45.20}\times x$.}
%\label{fig:benchvsoct}
%\end{figure}
A clear correlation can be seen for all channels with the exception of one chip that was measured at a high gain with the cosmics test bench, but has a more standard value with the muon measurement. This chip would appear at G$_{L}^{Oct} = 43$\,ADC~counts and G$_{L}^{Cosmics} = 260$ ADC~counts in Figure~\ref{fig:benchvsoct}. The correlation factor between the two datasets is found to be 75.8\,\%, discarding the outlier chips. None of the chips has deteriorated, confirming a stable and reliable behaviour of the chips with time.

\begin{figure}[h!]
\centerline{\includegraphics[width=0.85\columnwidth]{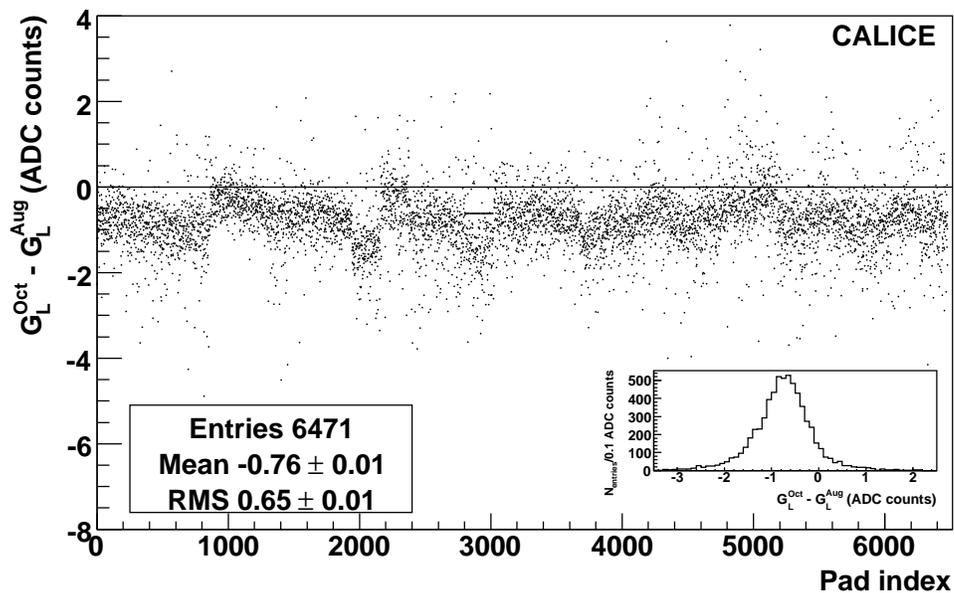}}
\caption{\sl Difference between October and August values, as a function of the pad index. The inset histogram displays the projection on the y-axis.}
\label{fig:diffaugoct}
\end{figure}

%To check the time dependence, the same method is applied to muon runs taken in August 2006 in similar conditions. 
The August sample results are plotted against the October sample results in Figure~\ref{fig:augvsoct}. A correlation coefficient of 93.2\,\% is found between both datasets. The difference between October and August values is plotted against the pad index in Figure~\ref{fig:diffaugoct}. The mean value of the difference for all channels is found to be $-0.76 \pm 0.01$\,ADC~counts with a spread of $0.67 \pm 0.01$\,ADC~counts.
 
The systematic shift of about 1.5\,\% of a MIP shown in Figure~\ref{fig:diffaugoct} is not yet understood, and is being investigated. If it is due to some timing offset in the triggers, such that the sample \& hold is offset with respect to the shaper peak, an additional correlated systematic uncertainty of about 1.5\,\% of a MIP would have to be considered on the energy scale, as the trigger used to record electron or pion beams could have different timing settings again.
Otherwise, if the shift is, for example, due to a different operation voltage of the VFE PCBs, then different calibration constants are required for the different periods of data taking.

%To check the time dependence, the sample is divided into groups of six runs (around 1.5 million events), and the distribution is fitted using the method described above. The resulting MPV is plotted against time in Figure~\ref{fig:calibvstime}, for nine channels of a chip of a randomly chosen PCB with a convergent fit (see Section~\ref{calibresults}).
%\begin{figure}[h!]
%\centerline{\includegraphics[width=0.95\columnwidth]{./sections/figures/CalConstOverTime_PCB10_C_ch2.eps}}
%\caption{Calibration constants versus time, for groups of six muon runs taken at CERN in October 2006. The different colour curves represent nine channels of the second chip of a randomly chosen PCB, situated in layer 9.}
%\label{fig:calibvstime}
%\end{figure}
%The channels don't have any common variations in time (so neither in temperature through no obvious day/night variations), suggesting the stability of the constants within $\pm 1.5$\,ADC~counts at maximum.

%XXX the previous value should be coming from the [mean +/- rms] values for all runs of the calib constant per group of run... XXX

% The same behaviour has been checked in several randomly chosen PCBs. The total sample will hence be used per channel to evaluate the calibration constants and their errors.
\subsubsection{Summary}

For 99.09\,\% of the channels, the calibration constants are obtained from a convergent fit with a statistical uncertainty of 0.5\,\% of a MIP and a systematic uncertainty of 0.4\,\% of a MIP. For 0.77\,\% of the channels, it is necessary to copy the calibration constant from a neighbouring pad. This results in a higher systematic uncertainty of 1.7\,\% of a MIP, coming from the standard deviation of the calibration constants obtained from a convergent fit per chip. The remaining 0.14\,\% of channels are declared dead.

The mean and standard deviation of the calibration constants for all channels are summarised in Table~\ref{tab:calconsts}, for the three measurement in time that have been made. The correlation coefficient between the October and the August (cosmics setup) samples is found to be 93.2\,\% (75.8\,\%) indicating a reasonable stability with time. The systematic shift of 1.5\,\% of a MIP found between the August and October samples however needs to be understood. Studies are ongoing by comparing to additional samples taken in 2007 at CERN.

\begin{table}[h!]
\begin{center}
\begin{tabular}{|c|c|c|c|} 
\cline{2-4}
\multicolumn{1}{c|}{} & Cosmics 2004-2006 & August 2006 & October 2006 \\
\hline
Mean (ADC~counts) & $161.54 \pm 0.37$ & $46.24 \pm 0.03$ & $45.48 \pm 0.03$ \\
\hline
RMS/Mean (ADC~counts) & $0.038 \pm 0.002$ & $ 0.0538 \pm 0.0004$ & $ 0.0543 \pm 0.0004$ \\
\hline
\end{tabular}
\caption{\sl Summary of the measurements of the calibration constants for all channels, made with three different samples. The measurements from the cosmics setup are taken with a different DAQ system, and with averages per chip.}
%\vspace{0.2cm}
\label{tab:calconsts}
\end{center}
\end{table}

%\section{Summary of issues}

\section*{Conclusions}
\label{conclu}

The CALICE ECAL Si-W prototype is a large scale prototype with nearly 10,000
channels when completed, of which 6,480 were studied in this document. The chosen technology, specifically high granularity together with
a compact structure, is compatible with the physics goals of an ILC, through
the use of a particle flow approach for the reconstruction of events. 
Since a final ILC detector could have a coverage of around 3,000\,m$^2$ of 
silicon, the wafer price must be kept low and this has been a primary aim of
the sensor design. Wafers have
been produced, and continue to be produced, by several manufacturers, both
to keep the price down and to test several fabrication technologies.
% Some features have been discovered in the design that need to be corrected (compensation of power supply instabilities at the chip level), or understood (crosstalk between guard ring and surrounding pads, and between pads) and effort is ongoing in these areas.

As expected when building and testing new prototypes, issues have been encountered. The mechanical design and DAQ have proven themselves very reliable. The affected parts in the design and test procedure were mainly the wafer production and the module design, and electronics issues with power supply lines and grounding.

The first problem encountered affected the module, once glued on the VFE PCB, with incompatibilities observed between the chemical passivation used by the manufacturers and the gluing process. This was solved with the producer, and has now proven reliable over period of years.

When the prototype was first put in beam, further issues were discovered at the module level. The first one concerns a capacitive coupling between the guard ring and the peripheral pads, discovered through the observation of so-called ``square events''. This is being investigated and modelled with simulation, in order to reduce the crosstalk in the next design. Methods are also being developed to identify these events reliably in the beam test data. The second issue, observed through the sudden shift of all the pedestal signals of a module, proportional to the amplitude of the signal recorded in the module (and hence to the energy of the incoming particle), is not yet understood. An intermittent bad contact between the Aluminium foil providing the ground and the module could explain such an effect but is hard to diagnose. An improved isolation and grounding system is being investigated.

The last issue encountered, observed through the sudden shift of all the pedestal signals of a VFE PCB by the same amount, has since been attributed to the non-isolation of the VFE PCB power supply lines. No working fix has been provided yet but the design is being corrected for the next prototype.

Large datasets have been acquired in beam tests at both DESY and CERN.
Pedestals were found with a remaining offset smaller than 0.2\,\% of a MIP, with channel-to-channel variations of $1.7 \pm 0.1$\,\% of a MIP, and run-to-run variations of $1.1 \pm 0.1$\,\% of a MIP in
electron and pion runs. However, in the muon runs used to calibrate 
the response of each channel, the channel and time variations are smaller than 0.4\,\% of a MIP, 
ensuring a negligible impact on the energy scale.

The noise is around 13\,\% of a MIP for all channels, with 20\,\% of the channels having time variation greater than 3\,\%. The impact on the energy resolution will be $0.13 \times \sqrt{N}$ in MIP units, with $N$ the number of hits recorded above threshold.

Overall, the calibration constants are found with an error of 0.5\,\% of a MIP. 
A systematic shift of less than 2\,\% of a MIP is found between values measured 
in the summer (August 2006) and in autumn (October 2006), indicating reasonable
stability. In addition, the comparison with cosmics test bench measurements
taken several years before the beam tests shows a good correlation between
the two sets of measurements, indicating
a stable and reliable behaviour of the wafers over periods of years.

\section*{Acknowledgements}

We would like to thank the technicians and the engineers who
contributed to the design and construction of the prototypes, including 
U.Cornett, G.Falley, K.Gadow, 
P.G\"{o}ttlicher, S.Karstensen and P.Smirnov. We also
gratefully acknowledge the DESY and CERN managements for their support and
hospitality, and their accelerator staff for the reliable and efficient
beam operation. 
%%%%%%%%%%%%%%  Add Fermilab later on %%%%%%%%%%%%%%%%%
We would like to thank the HEP group of the University of
Tsukuba for the loan of drift chambers for the DESY test beam.
The authors would like to thank the RIMST (Zelenograd) group for their
help and sensors manufacturing.
This work was supported by the 
Bundesministerium f\"{u}r Bildung und Forschung, Germany;
% by the DFG Excellence Cluster EXC153 of Germany ; 
by the Helmholtz-Nachwuchsgruppen grant VH-NG-206;
by the BMBF, grant no. 05HS6VH1;
by the Alexander von Humboldt Foundation (Research Award IV, RUS1066839 GSA);
% INTAS (Grant YSF150-00) ; 
by joint Helmholtz Foundation and RFBR grant HRJRG-002, Russian Agency for Atomic Energy, ISTC grant 3090;
by Russian Grants  SS-1329.2008.2 and RFBR0402/17307a
and by the Russian Ministry of Education and Science;
%by CICYT,Spain;
by CRI(MST) of MOST/KOSEF in Korea;
by the US Department of Energy and the US National Science
Foundation;
by the Ministry of Education, Youth and Sports of the Czech Republic
under the projects AV0 Z3407391, AV0 Z10100502, LC527  and by the
Grant Agency of the Czech Republic under the project 202/05/0653;  
and by the Science and Technology Facilities Council, UK.

\bibliography{bibentries}  

\begin{thebibliography}{10}
\expandafter\ifx\csname url\endcsname\relax
  \def\url#1{\texttt{#1}}\fi
\expandafter\ifx\csname urlprefix\endcsname\relax\def\urlprefix{URL }\fi

\bibitem{CALICE}
The {CALICE} collaboration,
  \url{http://polywww.in2p3.fr/activites/physique/flc/calice.html}.

\bibitem{ILC}
{T. Behnke \it{et al.} (ed.)}, Reference {D}esign {R}eport "{V}olume~4:
  {D}etectors" (2007), available at \url{http://lcdev.kek.jp/RDR}.

\bibitem{SiCAL}
{D. Bederede \it{et al.}}, "{SICAL}, a high precision silicon-tungsten
  luminosity calorimeter for {ALEPH}", Nucl. Instr. and Meth. A{\bf{365}}
  (1995) 117.

\bibitem{OPAL}
{G. Abbiendi \it{et al.} [OPAL Collaboration]}, "{P}recision luminosity for
  {Z0} lineshape measurements with a silicon-tungsten calorimeter", Eur. Phys.
  J. C{\bf{14}} (2000) 373.

\bibitem{SLD}
{S. Berridge \it{et al.}}, "{The small angle electromagnetic calorimeter at
  SLD: a 2m$^2$ application of silicon detector diodes}", IEEE Trans. Nucl.
  Sci. {\bf{36}} (1989) 339.

\bibitem{mark}
{M. A. Thomson}, {"Progress with Particle Flow Calorimetry"}, to appear in the
  proceedings of 2007 International Linear Collider Workshop (LCWS07 and
  ILC07), Hamburg, Germany, May-June 2007. e-Print: arXiv:0709.1360
  [physics.ins-det].

\bibitem{PFA}
{J. E. Brau \it{et al.}}, {"Calorimetry for the NLC detector"}, in the
  Proceedings of the 1996 DPF/DPB Summer Study on New Directions in High-energy
  Physics, Stanford 1997 (eConf C960625:DET077, 1996) 437.

\bibitem{epoxypreg}
{TEXIPREG CC120 ET443}, \\ \url{http://www.saatiamericas.com/Seal/index.html}.

\bibitem{glue}
{EPO}-{TEK} 4110 {Technical Data Sheet},
  \url{http://www.epotek.com/SSCDocs/datasheets/E4110.PDF}.

\bibitem{wacker}
\copyright {Wacker Chemie AG}, \\ \url{http://www.wacker.com}.

\bibitem{Moscow}
Moscow {S}tate {U}niversity, \\ \url{http://www.phys.msu.ru/eng/news/latest/}.

\bibitem{ONsemi}
{O{N} {S}emiconductor, {C}zech {R}epublic},
  \url{http://www.onsemi.com/PowerSolutions/content.do?id=1142}.

\bibitem{JTE}
{L. Evensen \it{et al.}}, "{G}uard ring design for high voltage operation of
  silicon detectors", Nucl. Instr. and Meth. A{\bf{337}} (1993) 44.

\bibitem{JFleuryIEEE}
{J.Fleury \it{et al.}}, "{SKIROC : A front-end chip to read out the imaging
  silicon-tungsten calorimeter for ILC}", IEEE-NSS, Honolulu, HI, USA {\bf{3}}
  (2007) 1847.

\bibitem{warren}
M. {W}arren, "{A VME Readout System} for the {CALICE Electromagnetic
  Calorimeter}", presented at {IEEE-NSS, Rome, Italy, Oct. 2004}.

\bibitem{fed}
{C. Foudas \it{et al.}}, "{The CMS Tracker readout Front End Driver}", IEEE
  Trans. Nucl. Sci. {\bf{52}} (2005) 2836.

\bibitem{sbs}
{SBS620 VMEbus-PCI bridge}, \\ \url{http://www.gefanuembedded.com}.

\bibitem{G4}
{S. Agostinelli \it{et al.}}, {"GEANT4: A simulation toolkit"}, Nucl. Instr.
  and Meth. A{\bf{506}} (2003) 250.

\bibitem{AHCAL}
F.~Sefkow, "{T}he scintillator {HCAL} testbeam prototype", {in the Proceedings
  of 2005 International Linear Collider Workshop (LCWS 2005), Stanford,
  California, March 2005,} \\
  \url{http://www.slac.stanford.edu/econf/C050318/papers/0913.PDF}.

\bibitem{TCMT}
D.~Chakraborty, {"The Tail-Catcher/Muon Tracker for the CALICE Test Beam"}, {in
  the Proceedings of 2005 International Linear Collider Workshop (LCWS 2005),
  Stanford, California, March 2005,} \\
  \url{http://www.slac.stanford.edu/econf/C050318/papers/0919.PDF}.

\bibitem{DCCern}
J.~Spanggaard, "{D}elay {W}ire {C}hambers - {A} {U}sers {G}uide", Tech. Rep.
  SL-Note-98-023-BI, CERN, Geneva (March 1998).

\bibitem{TMinuit}
"{MINUIT - Function Minimization and Error Analysis}", {CERN Program Library
  entry D506 (1998)},
  \url{http://wwwasdoc.web.cern.ch/wwwasdoc/minuit/minmain.html}.

\end{thebibliography}
\bibliographystyle{elsart-num}
%\bibliographystyle{plain}

%\appendix
%\input{sections/appendix.tex}
% \section{}
% \label{}

%\begin{thebibliography}{00}


% notes:

% \bibitem{label2}
% If there is a note, it should come last:
% \bibitem{label3} \note
% \end{subbibitems}

%\bibitem{}
%\input{biblio/bibiPrototype.tex}
%\input{biblio/bibiDaq.tex}
%\input{biblio/bibiBeamtests.tex}
%\input{biblio/bibiCommissioning.tex}

\end{document}